\documentclass[aip,floatfix,twocolumn,showpacs,preprintnumbers,amsmath,amssymb,superscriptaddress,longbibliography,reprint]{revtex4-1}
\usepackage{color}
\usepackage[usenames,dvipsnames,svgnames,table]{xcolor}
\usepackage[colorlinks=true,linkcolor=blue,urlcolor=blue,citecolor=blue]{hyperref}

\usepackage{mathtools}
\usepackage{graphicx}
\usepackage{dcolumn}
\usepackage{array}
\usepackage{lipsum}
\usepackage{bm}
\usepackage{subfigure}
\usepackage{amssymb}
\usepackage{multirow}
\usepackage{tabularx}
\usepackage{amsmath}
\usepackage{braket}

\newcommand{\br}{\bm{r}}
\newcommand{\bq}{\bm{q}}

\newcommand{\xc}{_\mathrm{{\scriptscriptstyle xc}}}

\renewcommand{\vec}[1]{\mathbf{#1}}

\begin{document}

\title{The Relevance of Electronic Perturbations in the Warm Dense Electron Gas}
\author{Zhandos Moldabekov}
\affiliation{Center for Advanced Systems Understanding (CASUS), D-02826 G\"orlitz, Germany}
\affiliation{Helmholtz-Zentrum Dresden-Rossendorf (HZDR), D-01328 Dresden, Germany}

\author{Tobias Dornheim}
\affiliation{Center for Advanced Systems Understanding (CASUS), D-02826 G\"orlitz, Germany}
\affiliation{Helmholtz-Zentrum Dresden-Rossendorf (HZDR), D-01328 Dresden, Germany}

\author{Maximilian B\"ohme}
\affiliation{Center for Advanced Systems Understanding (CASUS), D-02826 G\"orlitz, Germany}
\affiliation{Helmholtz-Zentrum Dresden-Rossendorf (HZDR), D-01328 Dresden, Germany}
\affiliation{Technische  Universit\"at  Dresden,  D-01062  Dresden,  Germany}

\author{Jan Vorberger}
\affiliation{Helmholtz-Zentrum Dresden-Rossendorf (HZDR), D-01328 Dresden, Germany}

\author{Attila Cangi}
\email{a.cangi@hzdr.de}
\affiliation{Center for Advanced Systems Understanding (CASUS), D-02826 G\"orlitz, Germany}
\affiliation{Helmholtz-Zentrum Dresden-Rossendorf (HZDR), D-01328 Dresden, Germany}

\begin{abstract}
Warm dense matter (WDM) has emerged as one of the frontiers of both experimental and theoretical physics and is challenging traditional concepts of plasma, atomic, and condensed-matter physics.
While it has become common practice to model correlated electrons in WDM within the framework of Kohn-Sham density functional theory, quantitative benchmarks of exchange-correlation (XC) functionals under WDM conditions are yet incomplete. 
Here, we present the first assessment of common XC functionals against exact path-integral Monte Carlo calculations of the harmonically perturbed thermal electron gas. This system is directly related to the numerical modeling of X-Ray scattering experiments on warm dense samples.  
Our assessment yields the parameter space where common XC functionals are applicable. 
More importantly, we pinpoint where the tested XC functionals fail when perturbations on the electronic structure are imposed. We indicate the lack of XC functionals that take into account the needs of WDM physics in terms of perturbed electronic structures. 
\end{abstract}

\maketitle

\section{Introduction}
Understanding transient states in warm dense matter (WDM) is one of the grand challenges of plasma physics that is currently being tackled in a number of experimental facilities~\cite{MBRKA09, tschentscher_photon_2017, MacDonald_POP21,hoffmann_cpp_18}.
In these experiments, WDM is generated, for example, due to laser-induced shock compression~\cite{Fortov_book, graziani-book, falk_wdm}. At the foundational level, probing WDM facilitates a better understanding of astrophysical objects such as planetary interiors~\cite{AGP99,NH04, Militzer_2008, manuel, NFKR11, KKNF12} and stars~\cite{CBFS00, saumon1, Daligault_2009, Haensel}. 
Furthermore, understanding WDM has great technological potential as it occurs during the fuel compression processes in inertial confinement fusion~\cite{hu_ICF}. Finally, exploring novel materials properties by driving matter through the WDM regime is an active area of research~\cite{doe-report-17, TRB08,Brongersma2015, KSM07}.    

The interpretation of WDM experiments relies on a strong interplay with theory and simulation, because typical parameters like density and temperature cannot be inferred solely from the experimental data.
The use of quantum Monte Carlo (QMC) techniques has recently paved the way for determining the exact properties of interacting electrons in the uniform electrons gas under WDM conditions~\cite{dornheim_physrep_18, dornheim_jcp_19-nn, dornheim_dynamic, dornheim_prl_18, dornheim_prl_20}. However, this method is computationally expensive and does not take into account the coupling of the electrons to the ions explicitly.
Kohn-Sham density functional theory (KS-DFT)\cite{KS65} has therefore emerged as the standard method to model the electronic structure in WDM, because it includes both electron-electron and electron-ion interactions.
Due to its balance of accuracy and computational efficiency~\cite{doi:10.1146/annurev-physchem-040214-121420}, KS-DFT enables calculating materials properties such as structural and electronic transport properties under the conditions relevant to the WDM community~\cite{ISI:000177971300006, ISI:000185240600035, ISI:000227459400099, PhysRevB.77.184201, PhysRevLett.102.115701, PhysRevLett.108.091102, KDBLCSMR15}.

Here, the central quantity is the KS potential -- a mean-field potential that mimics the electron-electron interaction, in principle, exactly. The key ingredient to the KS potential is the exchange-correlation (XC) functional. In practice, the XC functional needs to be approximated. From a historical perspective, the development of XC approximations has focused on the electronic structure of molecules and solids under ambient conditions~\cite{doi:10.1063/1.4704546}. These are commonly ranked in terms of increasing accuracy and computational cost on the so-called Jacob's ladder~\cite{doi:10.1063/1.1390175}. Starting at the lowest rung with the local density approximation (LDA)~\cite{KS65}, an array of XC approximations has been developed including, for instance, the generalized gradient approximation (GGA)~\cite{PhysRevA.38.3098, PBE}, the meta-GGA~\cite{PhysRevLett.91.146401, SCAN}, and hybrid functionals~\cite{doi:10.1063/1.464913, doi:10.1063/1.472933, doi:10.1063/1.1564060}.  
A key step most relevant for WDM modeling is the generalization of KS-DFT to finite temperature~\cite{mermin_65}. Based on this, several works have fleshed out the theoretical aspects of functional construction at finite temperature~\cite{PhysRevLett.107.163001, PhysRevB.84.125118, PhysRevB.93.195132, PhysRevB.93.205140}. Most recently, these have led to the construction of XC functionals that have an explicit temperature dependence \cite{PhysRevB.86.115101, PhysRevLett.112.076403, PhysRevB.88.161108, PhysRevB.88.195103, groth_prl, PhysRevB.101.245141}.  

Despite these efforts, functional development has not taken into account the needs of WDM physics  to a large extent beyond the inclusion of finite-temperature effects in the electronic structure~\cite{karasiev_importance,PhysRevE.103.013210}. Some exact conditions imposed for chemistry and solid-state physics are not relevant to WDM. 
For example, in contrast to solid-state physics, the surface energy has no significance in WDM, because the generated samples have no well defined surface, but rather transition to a plasma state~\cite{Olson, PhysRevE.98.063206, PhysRevE.68.036403}. Conversely, there are properties that are highly important in WDM, but have little value for solid-state physics. The most prominent example are perturbations imposed on the electronic structure with respect to the wave number $q$. An accurate description of such perturbations across a large range of $q$ is essential for WDM modeling. 
Accuracy is not only required in the long wavelength regime $q<2q_F$ which is dominated by collective electronic excitations, but also at large wave numbers $q>2q_F$ where so-called single-particle effects become important~\cite{hamann_prb_20, hamann_cpp_20, zhandos_pop18, moldabekov_pop15}. Achieving a high accuracy on a wide range of wave numbers is essential, e.g., for modeling X-Ray Thomson scattering (XRTS) experiments, which is a technique of paramount importance in WDM diagnostics~\cite{GGLR2003:demonstration}. 

In this paper, we therefore analyze the accuracy of common XC functionals in the WDM regime when perturbations on the electronic structure are imposed across a large range of wave numbers $q$.
To that end, we benchmark the results of KS-DFT calculations against path-integral QMC data which are considered exact within the given error bars.
Specifically, we employ four common XC functionals $-$ the LDA in the Perdew-Zunger parametrization~\cite{Perdew_LDA}, the GGAs PBE~\cite{PBE} and PBEsol~\cite{PBEsol}, and the meta-GGA SCAN~\cite{SCAN}.
We focus on the sensitivity of these XC functionals towards a perturbed electronic structure at a finite wave number $q$. 
Thereby, we uncover failures of these XC approximations. 
Our result strongly highlights the need for novel XC functionals that remain accurate under a perturbed electronic structure that is typically present in WDM. 
Furthermore, to our knowledge, this assessment of KS-DFT against QMC data with respect to electronic density perturbations in the WDM regime has not been performed yet.
While recent developments introduced above have addressed the explicit temperature dependence of XC functionals, they do not account for the physics that stems from perturbations at finite $q$. In our analysis, we therefore separate the influence of finite temperature in the electronic states from the effect of perturbations in $q$. We follow the common practice of using the listed ground-state XC approximations and including only the implicit temperature dependence of the electronic structure in terms of a Fermi-Dirac occupation of the KS states.

The paper is organized as follows: we introduce the theoretical aspects and the simulation methods in Section~\ref{s:theory}; we present our results on benchmarking common XC approximations in Section~\ref{s:results}; we address the issue of finite size effects in Section~\ref{s:finite_size}; we assess the performance of the aforementioned XC functionals in terms of the total energy in Section~\ref{s:energy}; we provide conclusions and an outlook on future perspectives in Section~\ref{s:end}. We provide methodological details on KS-DFT and path-integral Monte Carlo (PIMC) in the Appendix.

\section{Theory and Simulation Methods}\label{s:theory}
We begin our assessment with the following Hamiltonian that imposes electronic perturbations on a warm dense uniform electron gas (UEG):
\begin{align}\label{eq:H}
\hat{H} = \hat{H}_{\rm UEG} +  \sum_{k=1}^N \sum_{i=1}^{N_p} 2\, A\cos(\hat{\br}_k \cdot \hat{\bq}_i),
\end{align}
where $\hat{H}_{\rm UEG}$ denotes the standard Hamiltonian of the UEG~\cite{loos,quantum_theory,review}, $N$ the number of electrons, $N_p$ the number of external harmonic perturbations, $A$ their amplitude, and $\vec q_i$ the wave vector of the perturbations. In this work, we consider both one harmonic perturbation $N_p=1$ and the combination of two harmonic perturbations $N_p=2$. Note that we work within atomic units throughout.  

It has been shown that the electronic states described by Eq.~(\ref{eq:H}) are generated in WDM experiments~\cite{PhysRevLett.125.085001}. Furthermore, Eq.~(\ref{eq:H}) has been used to rigorously examine various fundamental physical properties of the electronic structure in WDM, such as the local field correction (LFC)~\cite{PhysRevLett.75.689, PhysRevLett.69.1837, dornheim_pre17, groth_jcp17} and the non-linear response~\cite{PhysRevLett.125.085001, PhysRevLett.125.235001, dornheim2021density} that are used to describe XRTS signals. Moreover, it turns out that the UEG model provides an excellent description of many electronic properties in WDM, because the electron-ion coupling can be relatively weak~\cite{PhysRevLett.125.235001, CPP2015, GRABOWSKI2020100905}.

The Hamiltonian in Eq.~(\ref{eq:H}) is a convenient device to control the degree of inhomogeneity and the wave number of the imposed perturbations on the UEG by tuning the parameters $A$ and $\vec q$, respectively. 
We analyze the performance of XC functionals with respect to electronic density perturbations by considering an amplitude range $0.02\leq A\leq 5$. This range covers regimes from weak density perturbations with $\delta n/n_0\ll1$ to strongly inhomogeneous electronic systems with $\delta n/n_0\gg 1$, where $\delta n=n-n_0$. 
Additionally, we consider perturbations imposed by tuning the wave number in the range $0.843~q_F \leq q_i\leq 5.9~q_F$. Thereby, we assess the performance of XC functionals at different length scales ranging from the long wavelength regime defined by collective behavior $q<q_F$~\cite{hamann_prb_20, hamann_cpp_20} to the scale defined by the single-particle limit $q\gg q_F$~\cite{zhandos_pop18, dornheim_jcp_19-nn}. This covers the entire range of wave numbers relevant to WDM generated in experiments.      

WDM conditions prevail when we consider matter at solid density and a temperature $T \sim T_F$ \cite{bonitz_pop_08, moldabekov_pre_18}, where $T_F=E_F/k_B$ denotes the Fermi temperature defined in terms of the Fermi energy $E_F$ and the Boltzmann constant $k_B$. In our analysis, we consider the densities $r_s=2$ and $r_s=6$ and a temperature $T=T_F$, where $r_s=a/a_B$ defines the number density of electrons which is given as the ratio between the Wigner-Seitz radius $a$ and the first Bohr radius $a_B$. At $T\lesssim T_F$, the parameter $r_s$ also characterizes electronic non-ideality~\cite{ott_epjd18, Moldabekov_cpp_21}. Therefore, it is used also as a coupling parameter.
These parameters are typically encountered in experiments with laser-driven and  shock-compressed solid targets~\cite{LANDEN2001465, PhysRevLett.98.065002, PhysRevLett.109.065002}. 

The KS-DFT calculations were performed with the GPAW code~\cite{GPAW1, GPAW2, ase-paper, ase-paper2}, which is a real-space implementation of the projector augmented-wave method. Details of the KS-DFT simulation parameters such as the number of $\vec k$-points and number of bands are given in the Appendix.

We provide unassailable data for benchmarking our KS-DFT results by carrying out direct path-integral QMC calculations based on Eq.~(\ref{eq:H}) without any restrictions on the nodal structure of the thermal density matrix. Therefore, the calculations are computationally expensive due to the fermion sign problem~\cite{dornheim_sign_problem, dornheim2021fermion}, but \emph{exact} within the given Monte Carlo error bars. The corresponding simulation details, such as the number of imaginary-time propagators, can be found in the Appendix~\ref{sec:app2}.

\begin{figure}
\center
\includegraphics{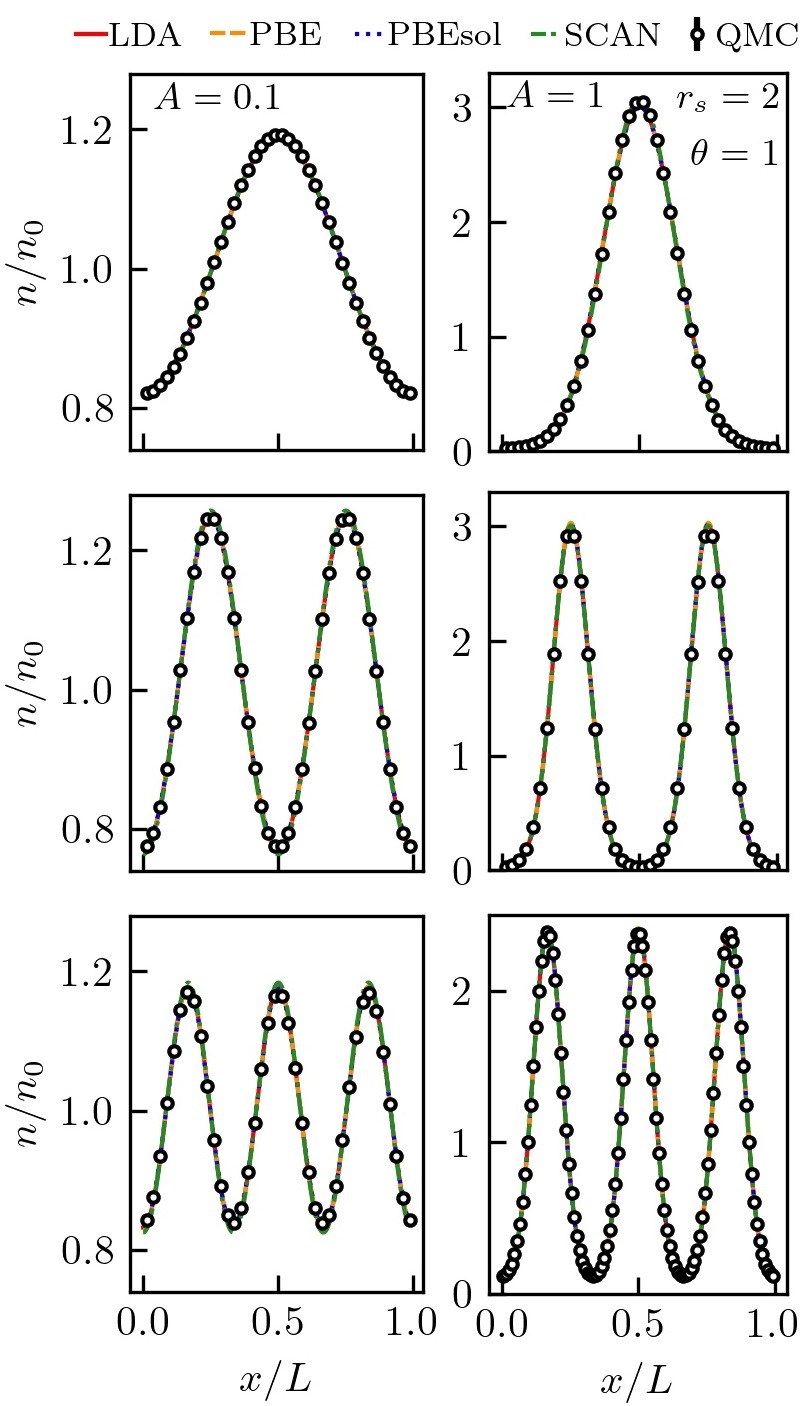}
\caption{ \label{fig:den_rs2} The electronic density distribution along the perturbation direction for two different amplitudes $A$ and increasing wave number $q$ at $r_s=2$ and $T=T_F$.
}
\end{figure} 
\begin{figure*}
\center
\includegraphics{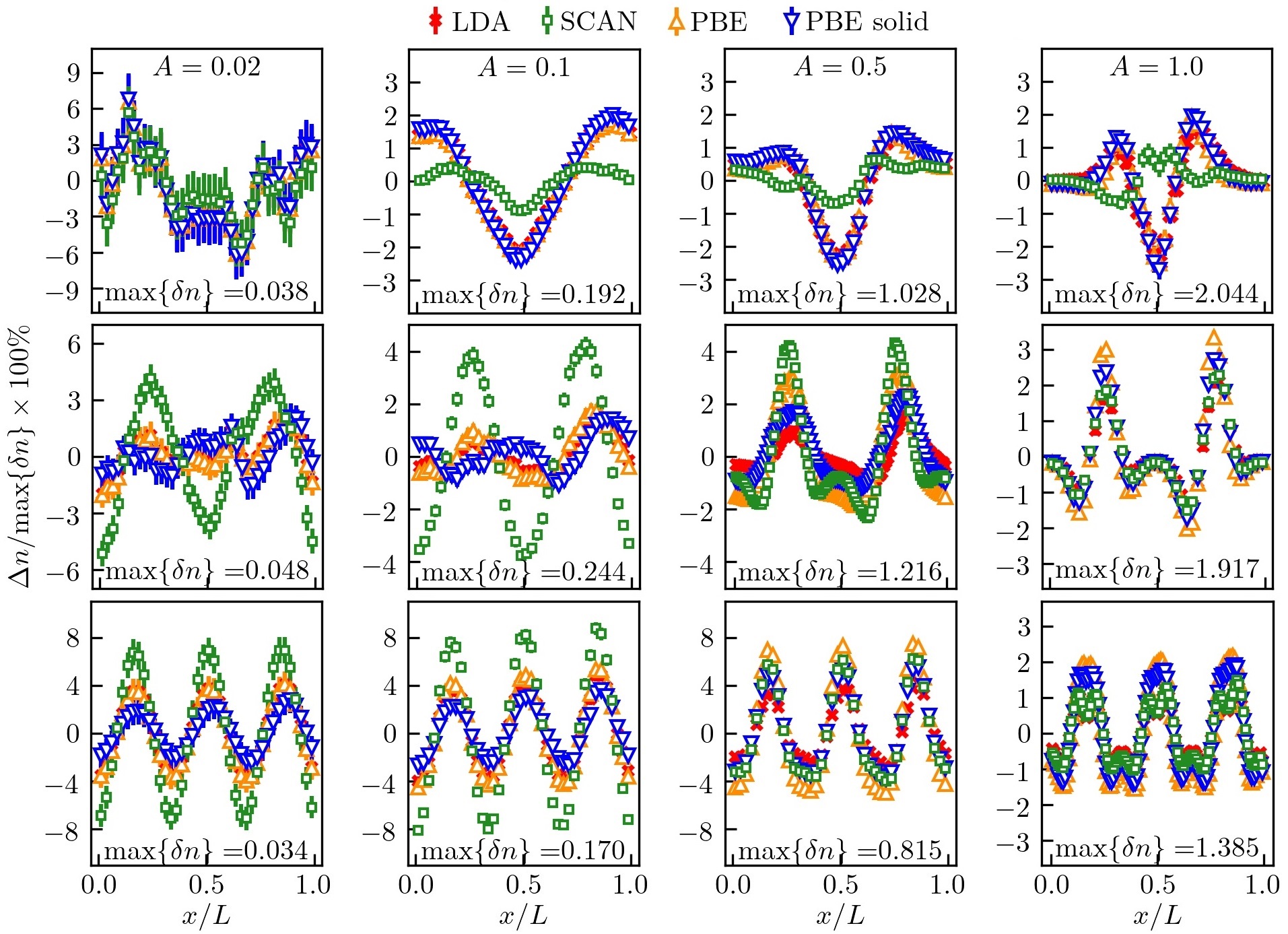}
\caption{\label{fig:den_err1}
Relative deviation in the density $\Delta n/{\rm max}\{\delta n\} 100~\%$ between the KS-DFT data and the reference QMC data at $r_s=2$. Column-wise from left to right: $A=0.02,~0.1,~0.5$ and $A=1$. 
Row-wise from top to bottom: $q_1=q_{\rm min},~2 q_{\rm min}$, and $q_1=3 q_{\rm min}$.}
\end{figure*}

\section{Results}\label{s:results}

\subsection{Single Harmonic Perturbation}
Let us begin our analysis by considering the case of a single harmonic perturbation, i.e., $N_p=1$ in Eq.~(\ref{eq:H}).
First, we investigate the electronic density distribution at a density (coupling) parameter $r_s=2$ for a range of perturbation amplitudes and wave numbers. Then, we consider a stronger coupling regime at $r_s=6$. These parameters are of particular interest in WDM experiments. 

\subsubsection{Metallic Density, $r_s=2$}
The density distribution along the direction of the density perturbation is shown for two amplitudes $A=0.1$ and $A=1$ and increasing wave number in Fig.~\ref{fig:den_rs2}.
The top panel displays the results for $q_1=q_{\rm min}=0.843~q_F $, the middle panel for $q_1=2 q_{\rm min}$, and the bottom panel for $q_1=3 q_{\rm min}$. On the scale of the total density, all considered XC functionals are in overall agreement with each other and with the QMC data.

Let us now consider the actual deviation of the KS-DFT data from the reference QMC data in closer detail.
To that end, we consider the relative density deviation $\Delta n/{\rm max}\{\delta n\}$ between the KS-DFT data and the reference QMC data, where ${\rm max}\{\delta n\}$ is the maximum deviation of the QMC data from the mean density. We use ${\rm max}\{\delta n\}$ for the analysis of the KS-DFT results because the physically important quantity in the case of a weak perturbation $\delta n/n_0\ll1$ is the deviation of the density from $n_0$ rather than the total density $n$ itself. Indeed, $\delta n$ defines the density response of the system which is a cornerstone of linear response theory describing all related physical properties of electrons in equilibrium. 
More specifically, when the external field defined via Eq.~(\ref{eq:H}) is weak, the density perturbation $\delta n=2A \chi(\vec q) \cos(\vec q\cdot \vec r)$ defines the static response function $\chi(\vec q)$. Therefore, $\Delta n/{\rm max}\{\delta n\}=\Delta \chi/\chi_{\rm QMC}$ measures the error in the response function.
On the other hand, to keep the present analysis general, we provide the values of ${\rm max}\{\delta n\}$ along with $\Delta n/{\rm max}\{\delta n\}$. This allows a simple conversion of data from $\Delta n/n_0$ or $\Delta n/n$.

In Fig.~\ref{fig:den_err1}, we show the relative difference between the KS-DFT data and the reference QMC data at $r_s=2$ for the amplitudes $A=0.02,~0.1,~0.5,~1$ (from left to right) and for the wave numbers $q_1=q_{\rm min},~2 q_{\rm min},~3 q_{\rm min}$ (from top to bottom). Additionally, the corresponding largest absolute value of the discrepancy is given in Table \ref{Table1}.

\begin{table*}[]
\caption{The performance of common XC functionals in terms of the relative density deviation $\Delta n/{\rm max}\{\delta n\} 100~\%$. A single harmonic perturbation at a fixed density $r_s=2$ (metallic density) and varying perturbation amplitude $0.02\leq A\leq 1$ and wave number $q_{\rm min}\leq q\leq 3 q_{\rm min}$ is considered, where $q_{\rm min}=0.843~q_F$. The largest absolute values of the deviation are listed in this table.}
\vspace{0.5cm}
\label{Table1}
\begin{tabular}{ c m{1em} ccc m{1em} ccc m{1em} ccc m{1em} ccc }
\hline   
\hline   
\multirow{2}{*}{} & & \multicolumn{3}{c}{$\bf A=0.02$} & & \multicolumn{3}{c}{$\bf A=0.1$} & & \multicolumn{3}{c}{$\bf A=0.5$} & & \multicolumn{3}{c}{$\bf A=1.0$} \\
\hline
& & $\bf q_{\rm min}$   & $\bf 2 q_{\rm min}$ & $\bf  3 q_{\rm min}$ 
& & $\bf q_{\rm min}$   & $\bf 2 q_{\rm min}$ & $\bf 3 q_{\rm min}$ 
& & $\bf q_{\rm min}$   & $\bf 2 q_{\rm min}$ & $\bf 3 q_{\rm min}$ 
& & $\bf q_{\rm min}$   & $\bf 2 q_{\rm min}$  & $\bf 3 q_{\rm min}$\\ 
 \hline 
{\bf LDA}    & & 6.59 & 2.01 & 4.07 & & 2.20 & 1.55 & 4.86 & & 2.37 & 1.57 & 4.22 & & 2.49 & 2.10 & 1.42\\ 
\hline
{\bf PBE}    & & 6.59 & 2.02 & 4.09 & & 2.20 & 1.72 & 5.34 & & 2.41 & 3.20 & 7.50 & & 2.52 & 3.36 & 2.18\\ 
\hline
{\bf PBEsol} & & 6.72 & 2.08 & 2.73 & & 2.40 & 1.40 & 5.60 & & 2.56 & 2.28 & 5.52 & & 2.72 & 2.71 & 1.92\\ 
\hline
{\bf SCAN}   & & 5.66 & 5.16 & 7.09 & & 0.95 & 4.24 & 8.75 & & 0.69 & 4.33 & 6.27 & & 0.86 & 2.29 & 1.49\\ 
\hline
\hline
\end{tabular}
\end{table*}

\begin{figure}
\center
\includegraphics{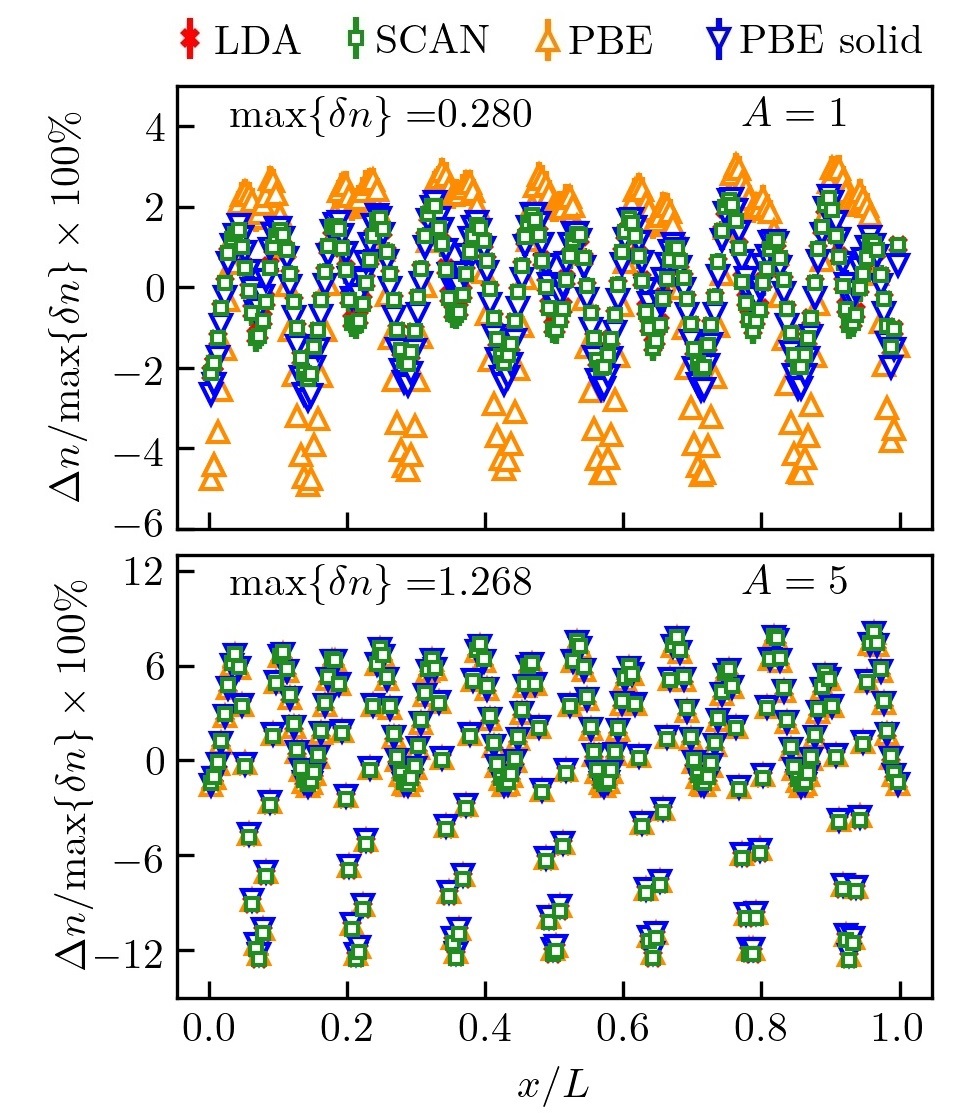}
\caption{\label{fig:den_err2}
Same as in Fig.~\ref{fig:den_err1}, but for $q_1=~7 q_{\rm min}$ at $A=1$ (top ) and $A=5$ (bottom).}
\end{figure} 

\begin{figure}
\center
\includegraphics{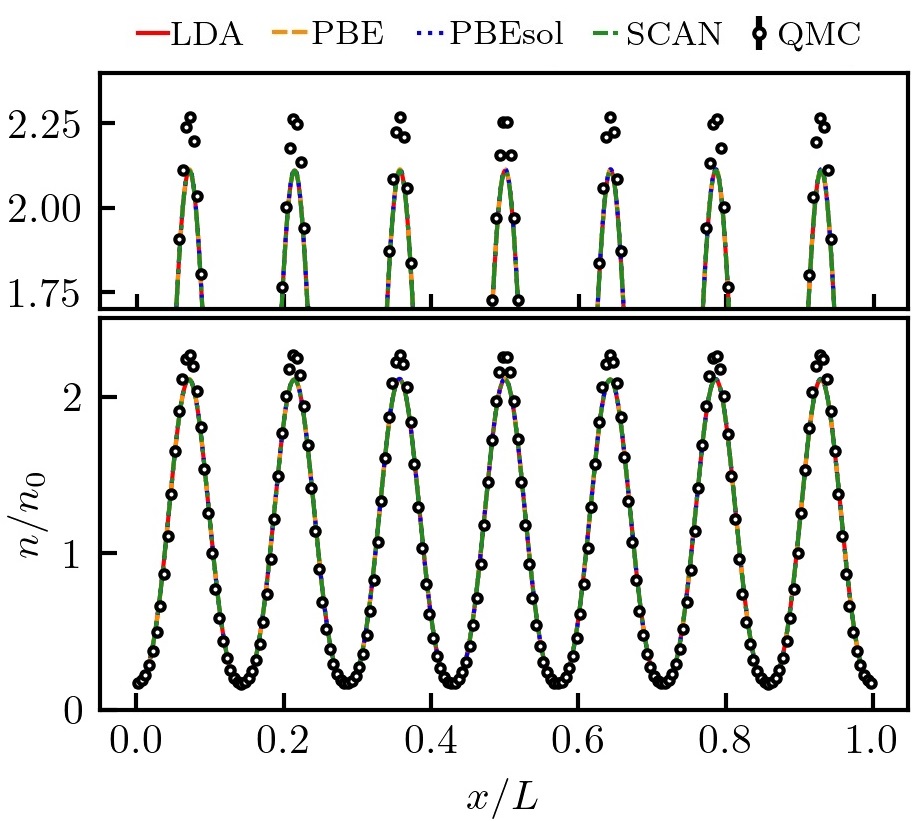}
\caption{\label{fig:density_A5}
The electronic density distribution along the perturbation direction for $A=5$ and $q_1=~7 q_{\rm min}$ at $r_s=2$.}
\end{figure} 

Our assessment of the results presented in Fig.~\ref{fig:den_err1} and Table~\ref{Table1} can be summarized as follows. 
At $A=0.02$ and $q_1=q_{\rm min}$, the results computed with the LDA, PBE, and PBEsol functionals have about the same level of accuracy. They are consistent with the QMC reference densities to within $7~\%$. With a maximum deviation of $5.66~\%$ from the QMC data, the SCAN functional yields slightly more accurate results. Note that here the statistical uncertainty of the QMC results is of the same order as the relative difference between the KS-DFT and QMC data (see the top left corner in Fig.~\ref{fig:den_err1}). Upon increasing the perturbation strength up to $A=1.0$, the relative error of the KS-DFT obtained from the LDA and GGA functionals remains less than $\leq 3~\%$, while the use of SCAN provides a remarkable accuracy better than $1~\%$. Therefore, the KS-DFT calculations using SCAN are virtually exact at $q_1=0.843~q_F $ for both weak and strong perturbations.
Contrarily, the accuracy provided by the SCAN functional is not maintained with an increase of the perturbation in terms of its wave number to $q_1=2q_{\rm min}$ and $q_1=3 q_{\rm min}$. Across the perturbation amplitdes $0.02\leq A\leq 0.5$ ($\delta n/n_0\leq 0.244$), the relative differences for calculations using SCAN are about $4~\%$ and $8~\%$. The LDA and PBEsol functionals provide an accuracy of about $2~\%$ ($5~\%$) at $q_1=2 q_{\rm min}$ ($q_1=3 q_{\rm min}$). The PBE functional provides a comparable accuracy to LDA and PBEsol at $q_1=2 q_{\rm min}$, but becomes less accurate at $q_1=3 q_{\rm min}$ and $A=0.5$ reaching an error of $7.5~\%$.
In the strong-perturbation regime ($A=1.0$ and ${\rm max}\{\delta n\}>1.3~n_0$), all functionals LDA, PBE, PBEsol functionals now provide about the same level of accuracy when compared to the QMC data. 
\textit{In summary, overall the LDA and PBEsol show a more robust performance compared to SCAN and PBE for all perturbation amplitudes and relatively small wave numbers at the typical mass density of metals.}

Next in Fig.~\ref{fig:den_err2}, we assess the performance of the XC functionals in the limit of large perturbation wave numbers by setting $q_1=7 q_{\rm min}=5.9~q_F$, but keeping the mass density metallic ($r_s=2$). We consider the perturbation amplitudes $A=1$ (with ${\rm max}\{\delta n\}= 0.28~n_0$) and $A=5$ (with ${\rm max}\{\delta n\}= 1.268~n_0$). 
At $A=1$, we find that for LDA, PBEsol, and SCAN, the relative difference between the KS-DFT data and the QMC data is less than $2.5~\%$, while the PBE functional yields an error of $5~\%$. 
When we increase the perturbation amplitude to $A=5$, the KS-DFT results from all considered functionals are in agreement with each other, but exhibit a strong disagreement with the QMC data of up to about $12~\%$. In this extreme regime, the deviation of the KS-DFT data from the QMC results is significant even on the scale of the total density $n$. This is illustrated in Fig.~\ref{fig:density_A5}.
\textit{In summary, all tested XC functionals fail dramatically in yielding an accurate electronic density for $q\gg q_F$ and ${\rm max}\{\delta n\}/n_0>1$ at metallic density.  We argue that a large number of other XC functionals that are derived from LDA, GGA, and meta-GGA classes are afflicted with the same limitation.}  

\begin{figure}
\center
\includegraphics{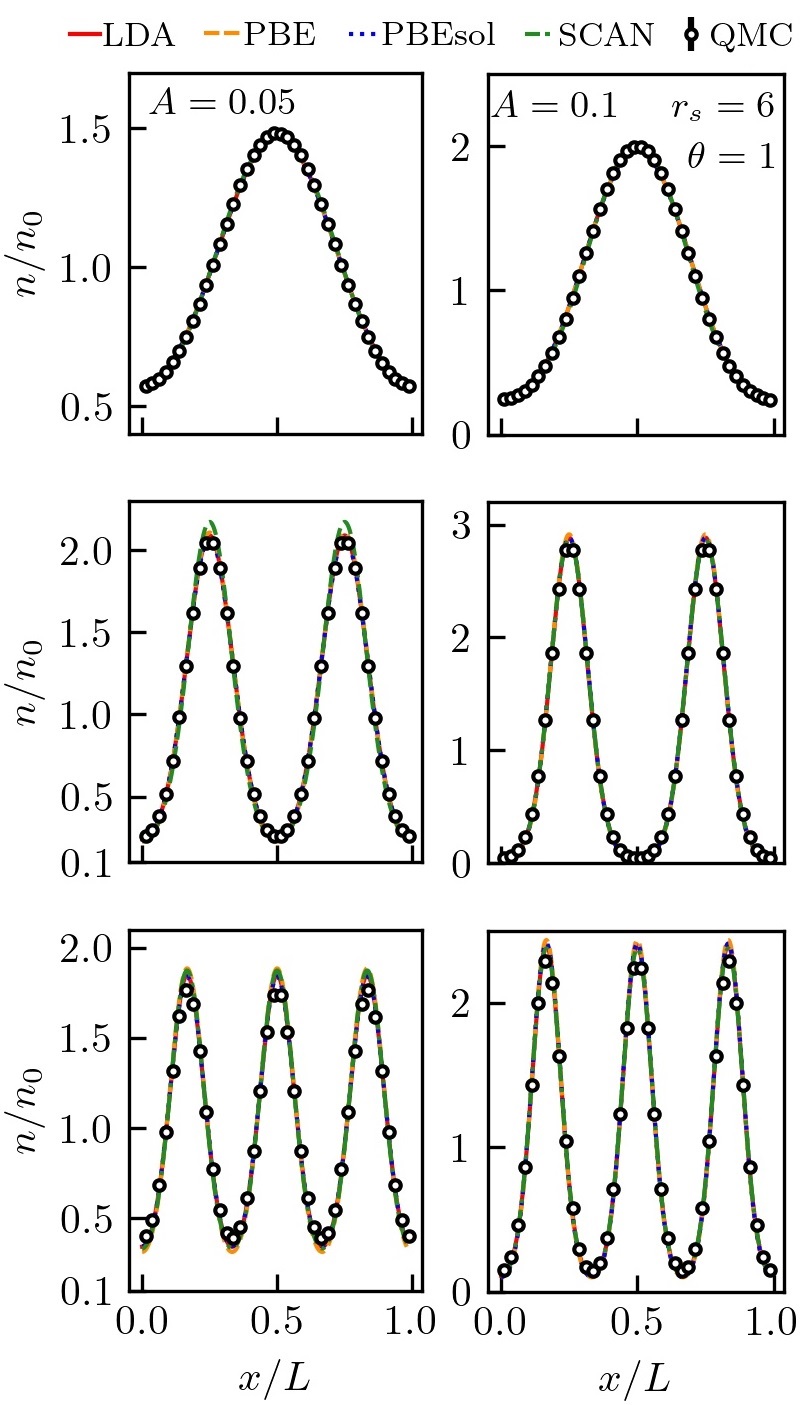}
\caption{\label{fig:density_rs6}
The electronic density distribution along the perturbation direction for two different amplitudes ($A=0.05$ and $A=0.1$) at $r_s=6$.}
\end{figure} 
\begin{figure*}
\center
\includegraphics{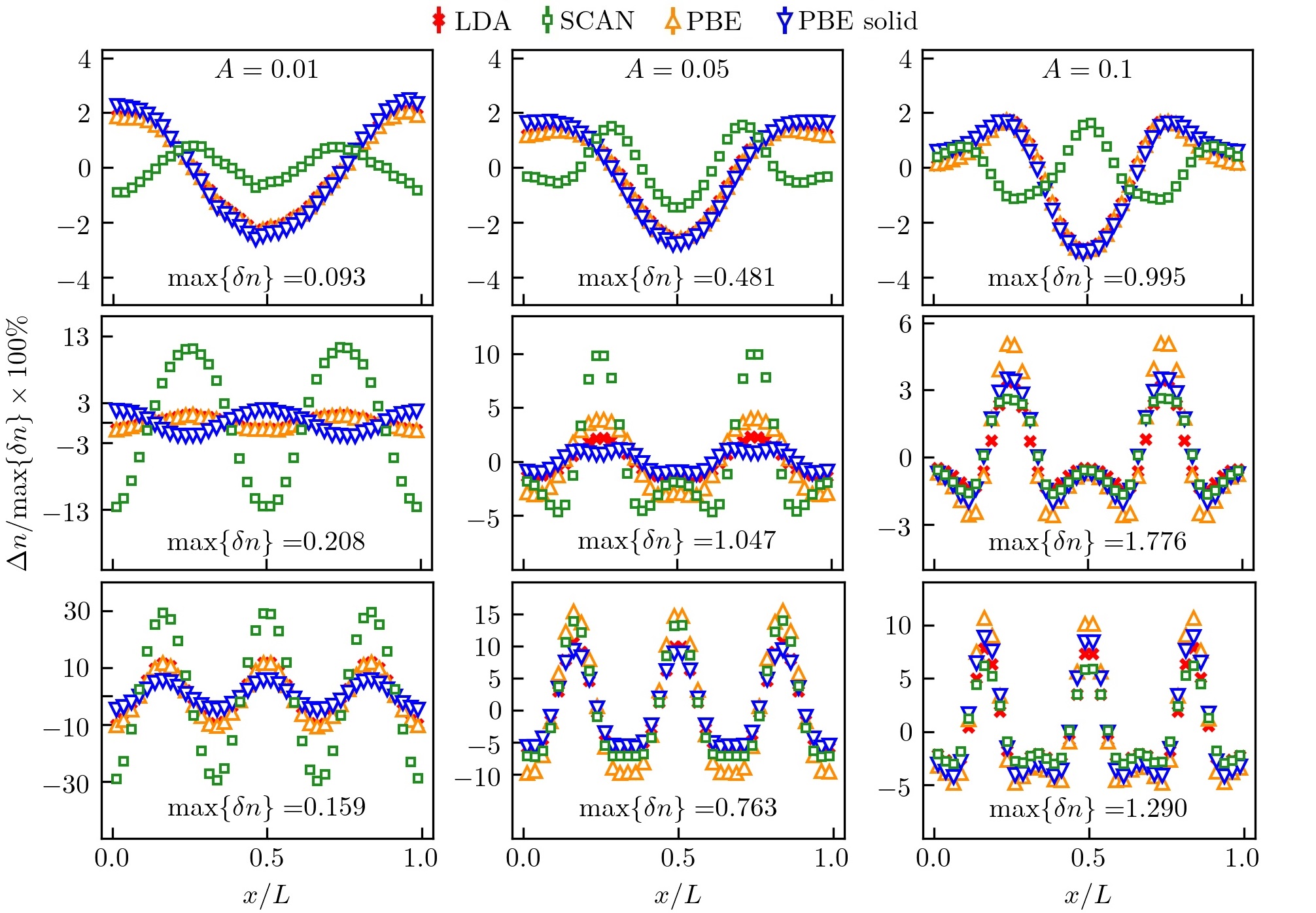}
\caption{\label{fig:den_err_rs6}
Relative deviation in the density $\Delta n/{\rm max}\{\delta n\} 100~\%$ between the KS-DFT data and the reference QMC data at $r_s=6$. Column-wise from left to right: $A=0.01,~0.05$ and $A=0.1$. 
Row-wise from top to bottom: $q_1=q_{\rm min},~2 q_{\rm min}$, and $q_1=3 q_{\rm min}$.}
\end{figure*} 

\subsubsection{Strong Coupling, $r_s=6$}
Next, we investigate strongly correlated electronic systems where $r_s=6$. Such low densities can be realized, for example, in evaporation experiments~\cite{Zastrau}.
From a theoretical perspective, these conditions are particularly challenging due to the substantial impact of electronic XC effects~\cite{low_density1,low_density2} on physical observables like the electrical conductivity or the density. 

Again, we consider weak as well as strong perturbation amplitudes corresponding to a density enhancement in the range $0.034~n_0\leq \delta n \leq 2.044~n_0$. 
The electronic density distribution along the direction of the perturbation is shown in Fig.~\ref{fig:density_rs6} for $q_1=q_{\rm min},~2 q_{\rm min}$, and $q_1=3 q_{\rm min}$ (from top to bottom) at $A=0.05$ (left) and $A=1.0$ (right).
The agreement between the KS-DFT data and QMC data is excellent when $q_1=q_{\rm min}$. With increasing wave number, the accuracy decreases. For example, the errors are larger than in the previous case where a metallic density is considered (see Fig.~\ref{fig:den_rs2}). 

We further delineate the relative difference between the KS-DFT data and the reference QMC results in Fig.~\ref{fig:den_err_rs6}.
There, the relative difference is illustrated for the amplitudes $A=0.01,~0.05$ and $A=0.1$ (from left to right) and the wave number $q_1=q_{\rm min},~2 q_{\rm min}$, and $q_1=3 q_{\rm min}$ (top to bottom) at $r_s=6$. The largest absolute values of the differences are listed in Table~\ref{Table2}.  

The main conclusions of our assessment in this parameters range are as follows.
At $q_1=q_{\rm min}$ and all perturbation amplitudes, the SCAN functional yields the most reliable results with an accuracy better than $1.64~\%$. The other tested  functionals provide an accuracy better than $3.13~\%$. 
In contrast to that, when we increase the wave number to $q_1=2 q_{\rm min}$, SCAN performs much worse with a maximum deviation of $12.58~\%$ at $A=0.01$ and $9.97~\%$ at $A=0.05$. Both the LDA and PBEsol functional yield an accuracy better than $3.5~\%$. Similarly, PBE yields an error of $4.02~\%$ and $5.08~\%$ at $A=0.05$ and $A=0.1$, respectively.  
However, irrespective of the wave number, the SCAN functional provides the most accurate results with the maximum deviation of $2.63~\%$ at a stronger perturbation amplitude $A=0.1$. 
Finally, increasing the wave number of the perturbation to $q_1=3 q_{\rm min}$ renders KS-DFT data less accurate with a maximal deviation in the range of about $8~\%$ and $12~\%$ (for $0.01\leq A\leq 0.1$) when LDA and PBE are used. The PBEsol functional is in better agreement with the QMC results showing a maximum deviation in the range between $5.63~\%$ and $9.45~\%$. The SCAN functional provides very low accuracy when the perturbation is weak($A=0.01$), with a deviation of almost $30~\%$. This improves in the case of strong perturbation $A=0.1$, where the maximum deviation is $6.2~\%$.    
\textit{In summary, the overall performance of the considered XC functionals is worse when strongly coupled electronic systems are considered. We observe failures of the functionals, in particular, when the wave number of the perturbation increases.} 

\begin{table*}[]
\caption{The performance of common XC functionals in terms of the relative density deviation $\Delta n/{\rm max}\{\delta n\} 100~\%$. A single harmonic perturbation at a fixed density $r_s=6$ (strong coupling) and varying perturbation amplitude $0.01\leq A\leq 0.1$ and wave number $q_{\rm min}\leq q\leq 3 q_{\rm min}$ is considered, where $q_{\rm min}=0.843~q_F$. The largest absolute values of the deviation are listed in this table.}
\vspace{0.5cm}
\label{Table2}
\begin{tabular}{ c m{1em} ccc m{1em} ccc m{1em} ccc}
\hline   
\hline   
\multirow{2}{*}{} & & \multicolumn{3}{c}{$\mathbf{A=0.01}$} & & \multicolumn{3}{c}{$\mathbf{A=0.05}$} & & \multicolumn{3}{c}{$\mathbf{A=0.1}$} \\
\hline
& & $\bf q_{\rm min}$   & $\bf 2 q_{\rm min}$ & $\bf  3 q_{\rm min}$ 
& & $\bf q_{\rm min}$   & $\bf 2 q_{\rm min}$ & $\bf 3 q_{\rm min}$ 
& & $\bf q_{\rm min}$   & $\bf 2 q_{\rm min}$ & $\bf 3 q_{\rm min}$\\ 
 \hline 
{\bf LDA}    & & 2.37 &  1.18 & 11.81 & & 2.66 & 2.29 & 10.60 & & 3.01 & 3.44 &  7.85\\ 
\hline
{\bf PBE}    & & 2.37 &  1.18 & 11.84 & & 2.61 & 4.02 & 15.57 & & 3.04 & 5.08 & 10.68\\ 
\hline
{\bf PBEsol} & & 2.63 &  2.04 &  5.63 & & 2.80 & 1.16 &  9.45 & & 3.13 & 3.48 &  8.86\\ 
\hline
{\bf SCAN}   & & 0.91 & 12.58 & 29.85 & & 1.55 & 9.97 & 14.02 & & 1.64 & 2.63 &  6.20\\ 
\hline
\hline
\end{tabular}
\end{table*}

\subsection{Double Harmonic Perturbation}
Furthermore, we assess the accuracy of the considered XC functionals when more complex perturbations are applied. To that end, we consider a double harmonic perturbation $N_p=2$ with $q_1=q_{\rm min}$ and $q_2=2 q_{\rm min}$. This allows us to check if the observed poor performance of the XC functionals manifests itself when the perturbations of different wave numbers are superimposed. Again, we consider both metallic densities ($r_s=2$) and strongly coupled systems ($r_s=6$). 
When $r_s=2$, we set the perturbation amplitude to $A=0.1$ resulting in a density perturbation of the order of $0.1~n_0$. When $r_s=6$, we set $A=0.01$ leading to a similar density perturbation of the order of $0.1~n_0$.

\begin{figure}
\center
\includegraphics{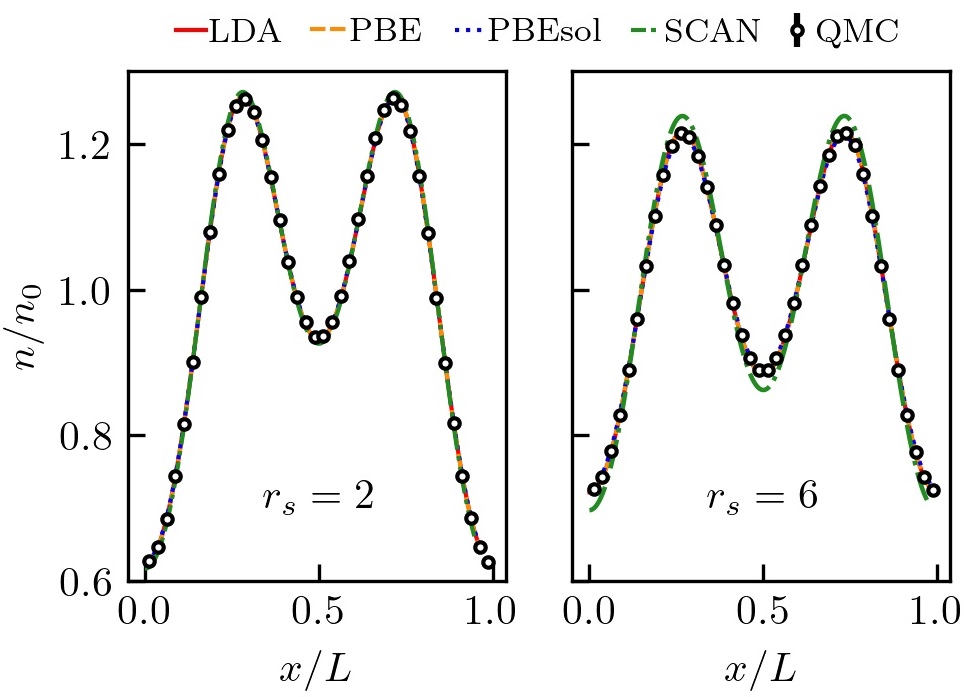}
\caption{\label{fig:density_two_A}
The density distribution along the perturbation direction for $r_s=2$ with $A=0.1$ (left) and for $r_s=6$ with $A=0.01$ (right). The wave numbers of the double harmonic perturbation in Eq.~(\ref{eq:H}) are $q_1=q_{\rm min}$ and $q_2=2 q_{\rm min}$. }
\end{figure}

\begin{figure}
\center
\includegraphics{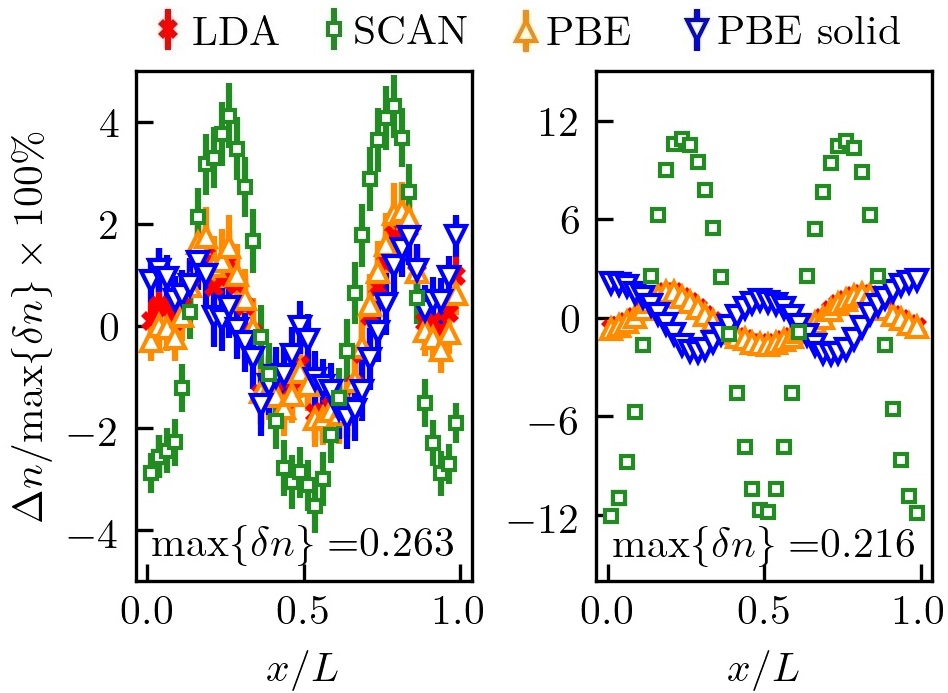}
\caption{\label{fig:density_err_two_A}
Relative deviation in the density $\Delta n/{\rm max }~\delta n \times 100~\%$ between the KS-DFT data and the reference QMC data for $r_s=2$ with $A=0.1$ (left) and for $r_s=6$ with $A=0.01$ (right). The wave numbers of the double harmonic perturbation in Eq.~(\ref{eq:H}) are $q_1=q_{\rm min}$ and $q_2=2 q_{\rm min}$.}
\end{figure}

The resulting electronic density distributions are shown in Fig.~\ref{fig:density_two_A}. At $r_s=2$, the KS-DFT results show good agreement with the QMC data when gauged on the scale of the total density. At $r_s=6$, the SCAN functional deviates from the QMC data significantly, while the results obtained from the other XC functionals are indistinguishable from the QMC data on this scale.  
We take a closer look at the performance of the various XC functionals in Fig.~\ref{fig:density_err_two_A}, where we use the relative deviation in the density $\Delta n/{\rm max}\{\delta n\}$ between the KS-DFT data and the reference QMC results. The corresponding largest absolute values of the differences are given in Table~\ref{Table3}. 

The results in Fig.~\ref{fig:density_err_two_A} and Table \ref{Table3} show that LDA, PBE, and PBEsol provide an accuracy better than $2.3~\%$ for both $r_s=2$ and $r_s=6$. Contrarily, the densities computed using SCAN have a maximum deviation of $4.31~\%$ and $12.05~\%$ at  $r_s=2$ and $r_s=6$, respectively. These numbers are similar to the deviations observed in the case of a single harmonic perturbation at the wave number $2 q_{\rm min}$. 
This provides a strong indication about the general applicability of the present findings to other systems, as any external potential can be expressed as a superposition of harmonic perturbations in reciprocal space.


\begin{table}[]
\caption{The largest absolute value of $\Delta n/{\rm max}\{\delta n\} 100~\%$ for $r_s=2$ with $A=0.1$  and for $r_s=6$ with $A=0.01$. The wave numbers of the double harmonic perturbation in Eq.~(\ref{eq:H}) are $q_1=q_{\rm min}$ and $q_2=2 q_{\rm min}$. }
\vspace{0.5cm}
\label{Table3}
\begin{tabular}{ c c c }
\hline
\hline
$\mathbf{r_s}$ & {\bf 2.0} & {\bf 6.0} \\
\hline 
{\bf LDA}      &  1.89   &  1.66   \\
\hline
{\bf PBE}      &  2.26   &  1.66   \\
\hline
{\bf PBEsol}   &  1.82   &  2.27   \\
\hline
{\bf SCAN}     &  4.31   &  12.05  \\ 
\hline
\hline
\end{tabular}
\end{table}


\section{Finite size effects}\label{s:finite_size}

\begin{figure}
\center
\includegraphics{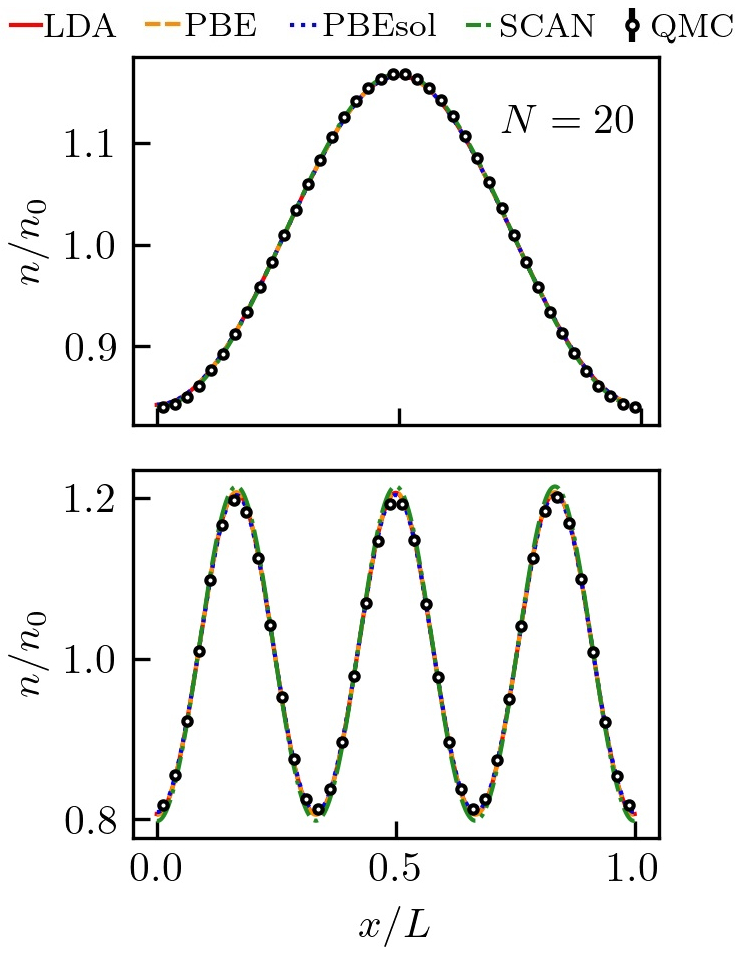}
\caption{\label{fig:den_N20}
The electronic density distribution along the perturbation direction for two different values of wave numbers $q=q_{\rm min}$ (top) and $q=3q_{\rm min}$ (bottom) at $A=0.1$ and $r_s=2$.}
\end{figure}

\begin{figure}
\center
\includegraphics{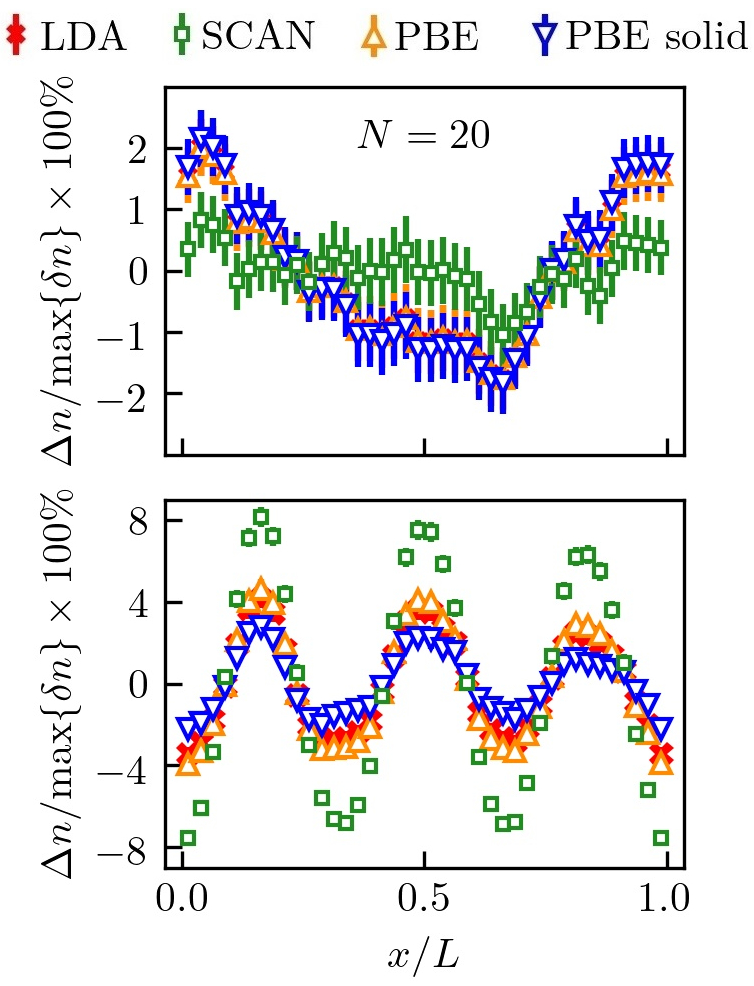}
\caption{\label{fig:dif_N20}
Relative deviation in the density $\Delta n/{\rm max}\{\delta n\} 100~\%$ between the KS-DFT data and the reference QMC data for  $q=q_{\rm min}$ (top) and $q=3q_{\rm min}$ (bottom) at $r_s=2$ and $A=0.1$.}
\end{figure} 

\begin{figure}
\center
\includegraphics{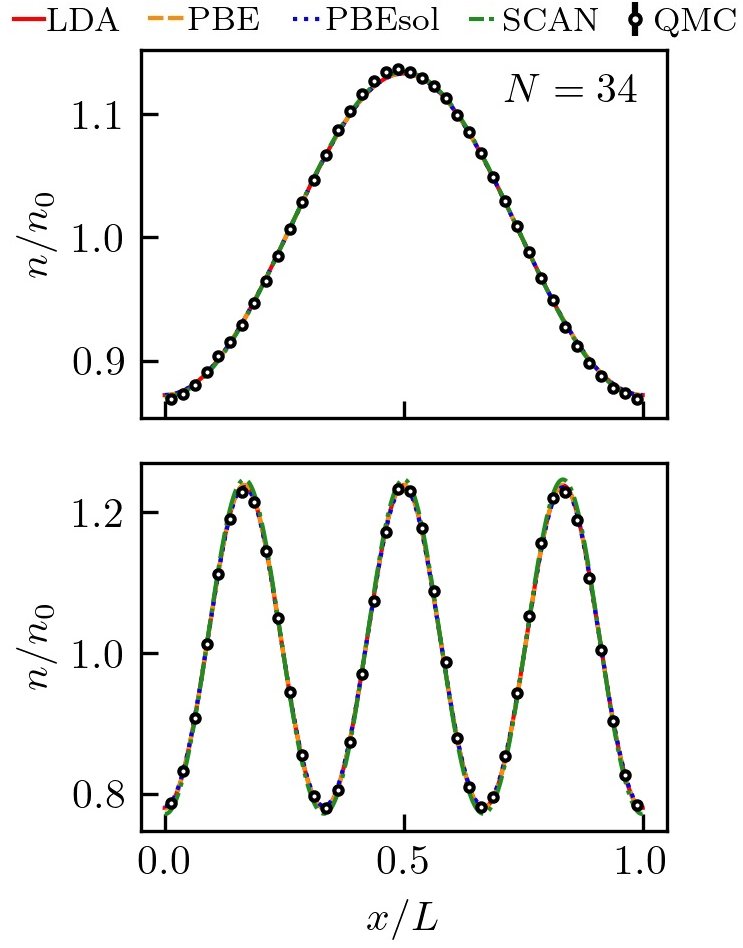}
\caption{\label{fig:den_N34}
The electronic density distribution along the perturbation direction for two different values of wave numbers $q=q_{\rm min}$ (top) and $q=3q_{\rm min}$ (bottom) at $A=0.1$ and $r_s=2$.}
\end{figure} 

\begin{figure}
\center
\includegraphics{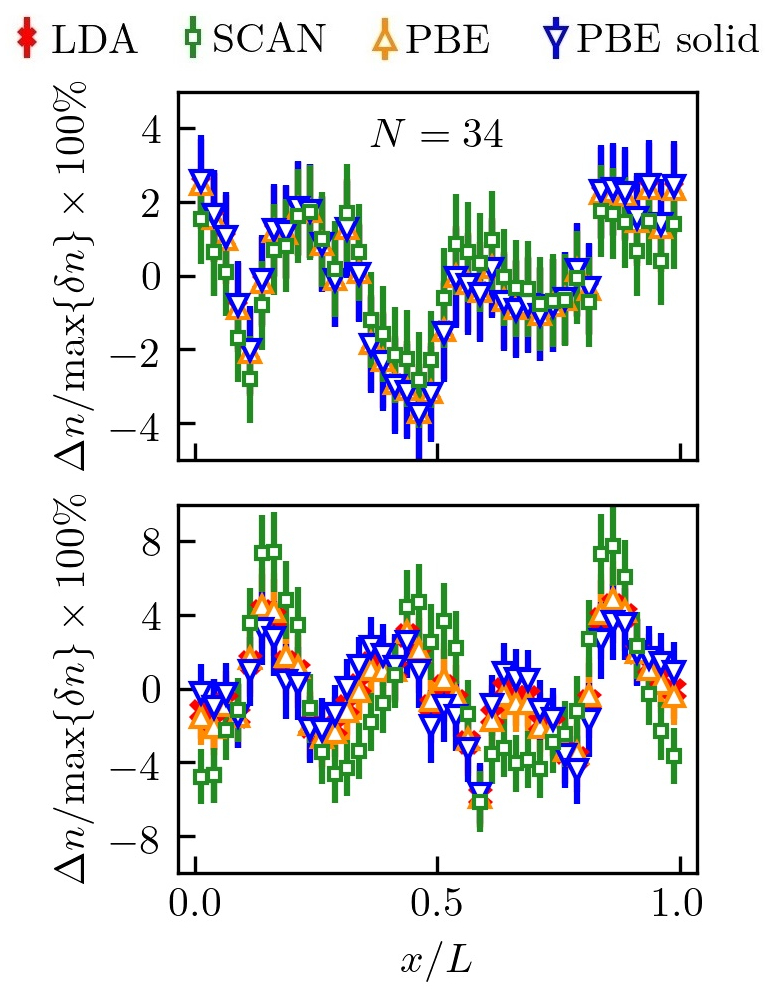}
\caption{\label{fig:dif_N34}
Relative deviation in the density $\Delta n/{\rm max}\{\delta n\} 100~\%$ between the KS-DFT data and the reference QMC data for  $q=q_{\rm min}$ (top) and $q=3q_{\rm min}$ (bottom) at $r_s=2$ and $A=0.1$.}
\end{figure} 

The results presented so far were computed for $14$ electrons in a simulation cell. Naturally, it is important to inquire whether the large deviations of the KS-DFT results from the QMC data observed at $2q_{\rm min}$ and $3q_{\rm min}$ compared to the case with $q_1=q_{\rm min}$ are indeed genuine results or an artifact due to finite size effects. 
To exclude this possibility, we consider $N=20$ and $N=34$ electrons in the simulation cell at $r_s=2$ and $A=0.1$. 
The density distribution along the direction of the density perturbation for the case with $20$ electrons is shown in Fig.~\ref{fig:den_N20}. The top panel displays the results for $q_1=q_{\rm min}=0.843~q_F $ and the bottom panel for $q_1=3 q_{\rm min}$. 
We observe that on the scale of the total density, the KS-DFT results are in agreement with the QMC data.
The corresponding relative difference between the KS-DFT results and the reference QMC data is displayed in Fig.~\ref{fig:dif_N20}, where the top panel is for $q_1=q_{\rm min}$ and the bottom panel is for $3 q_{\rm min}$. Fig.~\ref{fig:dif_N20} shows that at $q_1=q_{\rm min}$ the relative difference between KS-DFT results obtained using the SCAN functional and the QMC data is within the given uncertainty range, whereas it increases to up to $8\%$ with an increase in the wave number of the perturbation to $3q_{\rm min}$. The other considered XC functionals lead to slightly worse agreement at $q_1=q_{\rm min}$, with about $2\%$ maximum relative difference, and to much better results at $q_1=3q_{\rm min}$, with a maximum relative difference of about $4\%$. 

Next, we further increase the number of electrons in the simulation cell to $34$. The density distribution for $34$ particles along the direction of the density perturbation is shown in Fig.~\ref{fig:den_N34} at  $q_1=q_{\rm min}$ (top panel) and at $q_1=3q_{\rm min}$ (bottom panel).
Again, on the scale of the total density we observe agreement of KS-DFT results with the QMC data. The corresponding relative differences, shown in Fig.~\ref{fig:dif_N34}, confirm this assessment. At $q_1=q_{\rm min}$ (top panel) the KS-DFT results agree with the QMC data within the given numerical uncertainty. At $q_1=3q_{\rm min}$ (the bottom panel in Fig.~\ref{fig:dif_N34}), KS-DFT results obtained with the SCAN functional exhibit a relative deviation from QMC data of about  $8\%$. The other XC functionals yield a maximum relative difference of about $4\%$. Note the statistical uncertainty of the QMC data at $N=34$ is noticeably larger due to the infamous fermion sign problem. See Appendix for more details and Ref.~\onlinecite{dornheim_sign_problem} for a topical review of this issue. 

This analysis of the densities and relative differences between KS-DFT results and QMC data for $N=20$ and $N=34$ electrons clearly shows that our conclusions at $N=14$ electrons are not affected by finite size effects.

\section{Energy perturbation}\label{s:energy}

\begin{figure}
\center
\includegraphics{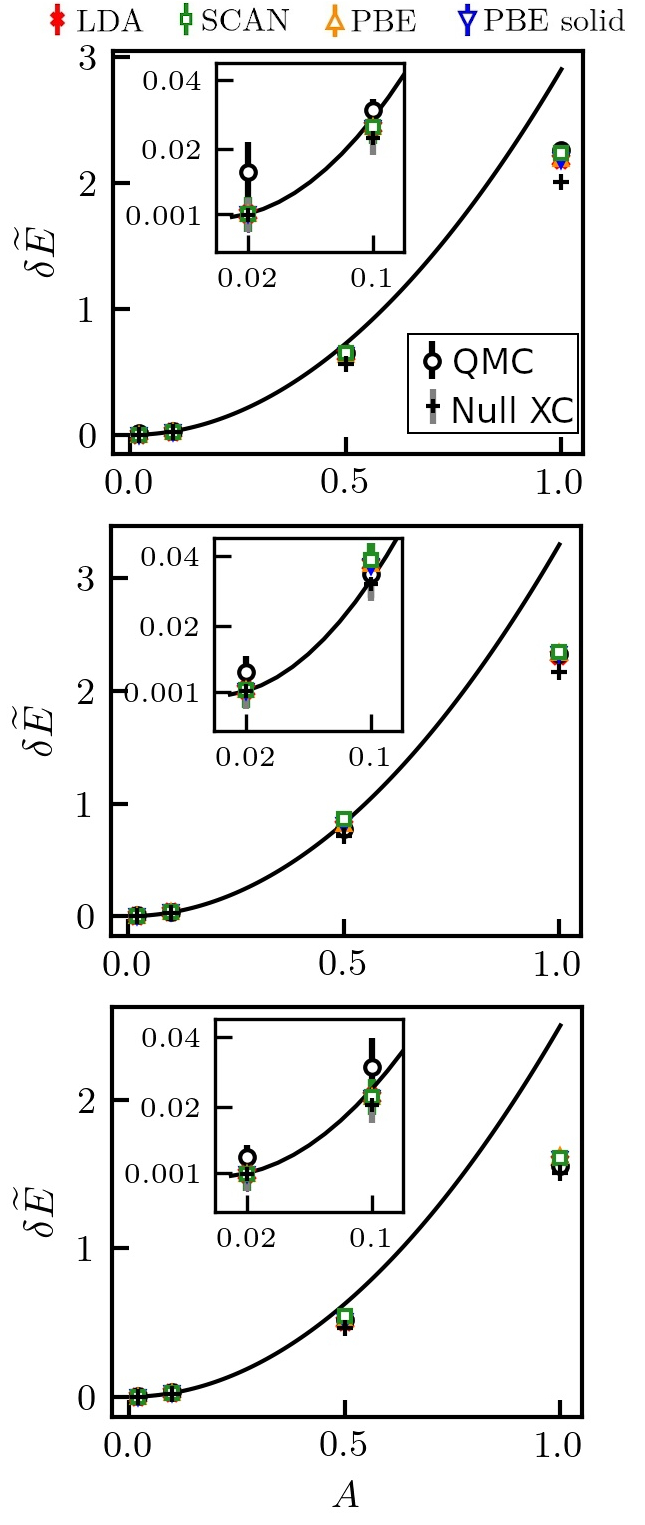}
\caption{\label{fig:energy}
Relative deviation in the total energy from UEG ($A=0$) due to external perturbation with amplitude $A$ (see Hamiltonian (\ref{eq:H})) at $r_s=2$. From top to bottom: $q_1=q_{\rm min},~2 q_{\rm min}$, and $q_1=3 q_{\rm min}$.}
\end{figure} 

\begin{figure}
\center
\includegraphics{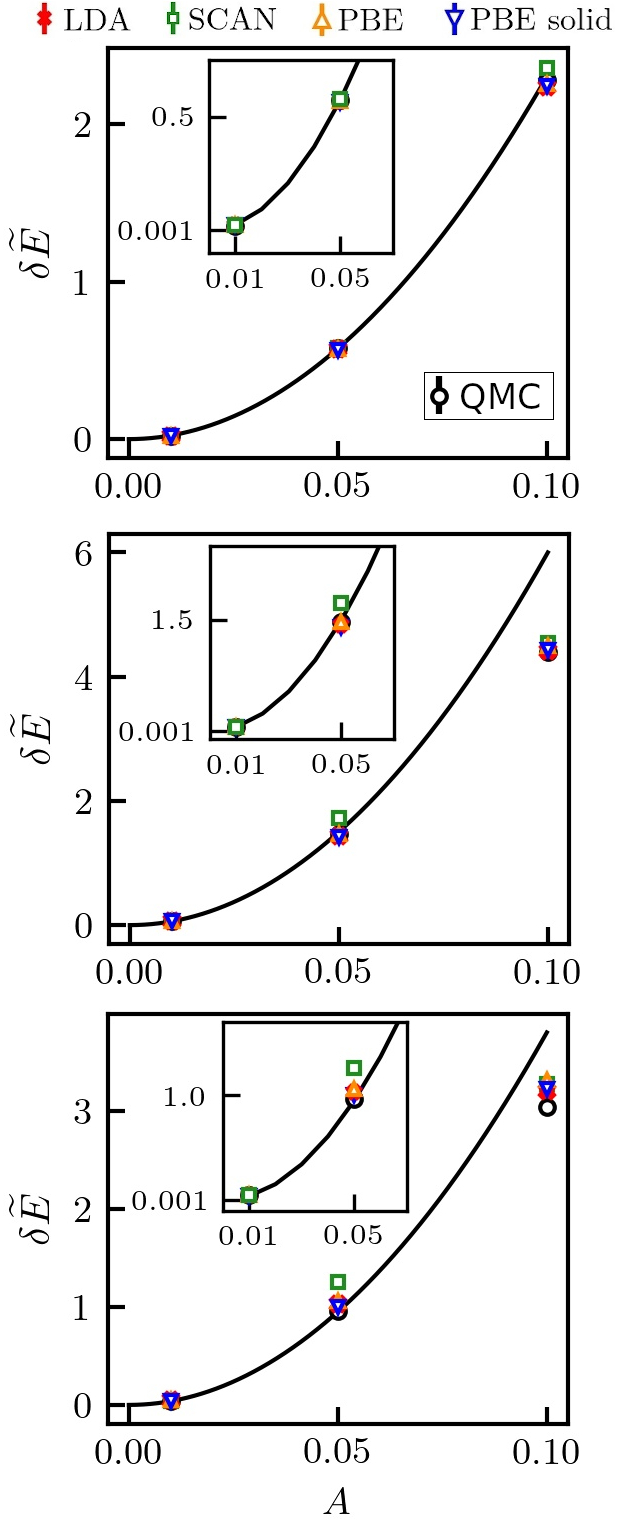}
\caption{\label{fig:energy_rs6}
Relative deviation in the total energy from UEG ($A=0$) due to external perturbation with amplitude $A$ (see Hamiltonian (\ref{eq:H})) at $r_s=6$. 
Row-wise from top to bottom: $q_1=q_{\rm min},~2 q_{\rm min}$, and $q_1=3 q_{\rm min}$.}
\end{figure} 

Finally, we analyze the correlation of the revealed inaccuracies in the densities with the total energy.
To that end, we focus on the change in the total energy compared to the unperturbed case. This is quantified by the reduced perturbation in the total energy
\begin{equation}\label{eq:dE}
    \delta \widetilde{E}= \frac{\left|E(A)-E(A=0)\right|}{\left|E_{\rm QMC}(A=0)\right|},
\end{equation}
where $E(A)$ is the total energy of electrons in the presence of the external perturbation with the amplitude $A$, $E(A=0)$ is the total energy of the unperturbed system, and $E^{\rm QMC}(A=0)$ is the total energy of the unperturbed system from QMC simulations.

The results for $\delta \widetilde{E}$ at $r_s=2$ and at  $r_s=6$ for different amplitudes and wave numbers are presented in Fig.~\ref{fig:energy} and Fig.~\ref{fig:energy_rs6}, respectively. In these figures we also plot $\delta \widetilde{E}\sim A^2$ using the data point at $A=0.1$ ($A=0.01$) for $r_s=2$ ($r_s=6$). This dependence is known to be exact in the limit of small $A$ in the ground state at $T\ll T_F$ and is referred to as the stiffness theorem \cite{giuliani2005quantum}. At finite temperature, the stiffness theorem applies to the total free energy. Figs.~\ref{fig:energy} and \ref{fig:energy_rs6} can be understood as an empirical indication of validity of stiffens theorem for the total energy at finite temperature. 
The deviation of the simulation data from the $A^2$ dependence is related to the emergence of the higher-order non-linear effects.

In Fig.~\ref{fig:energy}, in addition to KS-DFT results within LDA, PBE, PBEsol, and SCAN, we also show results computed using null XC functional, i.e., the electron-electron interaction approximated only using the mean Hartree field. First of all, in Fig.~\ref{fig:energy} we clearly see that neglecting the XC energy leads to less accurate results compared to the QMC data at amplitudes $A=0.1$, $A=0.5$ and $A=1.0$. At $A=0.02$, the change in the energy due to the external field is comparable with statistical uncertainty. This prevents us from quantifying the accuracy of the KS-DFT results for a reduced perturbation in the energy at that perturbation strength. 
At $A=0.1$, all KS-DFT data points, except the null XC functional case, are in agreement with the QMC data within the given statistical uncertainty. 
Furthermore, in Table \ref{Table_energy}, we provide the relative difference of the reduced perturbation in the total energy, $\Delta \widetilde E$, which is defined by

\begin{equation}\label{eq:DE}
    \Delta \widetilde E ~[\%]=\frac{|\delta \widetilde{E}_{\rm QMC}-\delta \widetilde{E}_{\rm DFT}|}{\delta \widetilde{E}_{\rm QMC}} \times 100\ .
\end{equation}

\begin{table}[]
\caption{The performance of common XC functionals in terms of the relative deviation in the total energy as defined by Eq.~(\ref{eq:DE}). A single harmonic perturbation at a fixed density $r_s=2$ and  perturbation amplitudes $A=0.5$ and $A=1.0$ and wave numbers $q_{\rm min}\leq q\leq 3 q_{\rm min}$ is considered, where $q_{\rm min}=0.843~q_F$. The presented data has about $1\%$ uncertainty.}
\vspace{0.5cm}
\label{Table_energy}
\begin{tabular}{ c m{1em} ccc m{1em} ccc}
\hline   
\hline   
\multirow{2}{*}{} & & \multicolumn{3}{c}{$\mathbf{A=0.5}$} & & \multicolumn{3}{c}{$\mathbf{A=1.0}$} \\
\hline
& & $\bf q_{\rm min}$   & $\bf 2 q_{\rm min}$ & $\bf  3 q_{\rm min}$ 
& & $\bf q_{\rm min}$   & $\bf 2 q_{\rm min}$ & $\bf 3 q_{\rm min}$\\ 
 \hline 
{\bf LDA}    & & 3.05  &  3.62  & 2.52 & & 3.63 & 0.70 & 2.29 \\ 
\hline
{\bf PBE}    & & 2.93&  5.84  & 0.75 & & 3.31 & 0.47 & 4.24 \\ 
\hline
{\bf PBEsol} & & 3.31 &  4.24 &  1.77 & &  3.62 & 1.36 &  3.23 \\ 
\hline
{\bf SCAN}   & & 0.54 & 9.92 & 5.56 & & 1.11 & 6.61 & 0.52 \\ 
\hline
{\bf Null XC}   & & 15.44 & 8.8 & 11.43 & & 12.84 & 7.47 & 2.98 \\ 
\hline
\hline
\end{tabular}
\end{table}

\begin{table}[]
\caption{The performance of common XC functionals in terms of the relative deviation in the total energy as defined by Eq.~(\ref{eq:DE}). A single harmonic perturbation at a fixed density $r_s=6$ and  perturbation amplitudes $A=0.05$ and $A=0.1$ and wave numbers $q_{\rm min}\leq q\leq 3 q_{\rm min}$ is considered, where $q_{\rm min}=0.843~q_F$. The presented data has about $1\%$ uncertainty. }
\vspace{0.5cm}
\label{Table_energy_rs6}
\begin{tabular}{ c m{1em} ccc m{1em} ccc}
\hline   
\hline   
\multirow{2}{*}{} & & \multicolumn{3}{c}{$\mathbf{A=0.05}$} & & \multicolumn{3}{c}{$\mathbf{A=0.1}$} \\
\hline
& & $\bf q_{\rm min}$   & $\bf 2 q_{\rm min}$ & $\bf  3 q_{\rm min}$ 
& & $\bf q_{\rm min}$   & $\bf 2 q_{\rm min}$ & $\bf 3 q_{\rm min}$\\ 
 \hline 
{\bf LDA}    & &  1.16 &  1.95 & 6.91& & 1.96 & 0.26 & 5.25\\ 
\hline
{\bf PBE}    & & 1.16 &  0.1  & 10.84& & 1.35 & 2.19 & 8.81 \\ 
\hline
{\bf PBEsol} & & 1.57 &  3.88 &  4.14 & &  1.80 & 0.55 &  5.80 \\ 
\hline
{\bf SCAN}   & & 0.43 & 17.80& 30.93 & & 3.07 & 3.15 & 7.42 \\ 
\hline
\hline
\end{tabular}
\end{table}

From Table \ref{Table_energy} we infer that in the case of null XC functional, $\Delta \widetilde E$ is about $15\%$ ($9\%$) and $13\%$ ($7\%$) for $A=0.5$ and $A=1.0$ at $q=q_{\rm min}$ ($q=2q_{\rm min}$). This large difference mainly stems from an inaccurate density due to neglecting the XC functional (see Appendix). This clearly demonstrates the importance of including the XC functional in the considered KS-DFT calculations.
Also, as shown in Table \ref{Table_energy}, the LDA, PBE, and PBEsol functionals show similar performance.  The fact that SCAN is more accurate for $\Delta \widetilde E$ at $q_1=q_{\rm min}$ and less accurate at $q_1=2q_{\rm min}$ correlates with the previous assessment of the densities.   
  
In the case of $r_s=6$ (Fig.~\ref{fig:energy_rs6}), the dependence of $\delta \widetilde{E}$ on $A$ is similar to that at $r_s=2$. At $A=0.01$, the KS-DFT results computed using different XC functionals is in agreement with the QMC data within the given statistical uncertainty. At $A=0.05$ and $A=0.1$, the SCAN results tend to overestimate $\delta \widetilde{E}$ with increasing wave number from $q_1=q_{\rm min}$ to $q_1=2q_{\rm min}$ and $q_1=3q_{\rm min}$. The LDA, PBE, and PBEsol results are very similar across all considered $A$ and $q_1$ values.

In Table~\ref{Table_energy_rs6} we list $\Delta \widetilde E$ as defined by Eq.~(\ref{eq:DE}) in order to better delineate the differences. Similar to $r_s=2$, also at $r_s=6$ the SCAN results are most accurate at $q_1=q_{\rm min}$. Contrarily, at $q_1=2q_{\rm min}$ and $q_1=3q_{\rm min}$, the largest deviation from the QMC data is exhibited by the SCAN results. This reflects the fact that there are more inaccuracies in the density when SCAN is used over the other XC functionals. The KS-DFT results for $\delta \widetilde{E}$ computed using LDA, PBE, and PBEsol have similar values at $q_1=q_{\rm min}$ and $q_1=2q_{\rm min}$. At $q_1=3q_{\rm min}$, the PBE results are less accurate by a few percent than the LDA and PBEsol results. Overall, by comparing data in Table \ref{Table2} and in Table \ref{Table_energy_rs6}, we observe that the variations in $\Delta \widetilde E$ reflect those in $\Delta n/{\rm max}\{\delta n\}$.

\section{Conclusions and Outlook}\label{s:end}

We benchmarked the performance of KS-DFT based on the LDA, PBE, PBEsol, and SCAN XC functionals against exact QMC data in the WDM regime. Our assessment revealed a set of conditions for the successful simulation of WDM with KS-DFT at QMC level accuracy. Our comparative analysis unambiguously demonstrates when the considered XC functionals fail to correctly describe the electronic density. 

We found that the KS-DFT results are sufficiently accurate for small wave numbers of the density perturbation, $q<q_F$. In particular, using the SCAN functional yields an excellent agreement with the QMC reference data. However, with increasing wave number $q>q_F$ (as tested for $q=1.686q_F$ and $q=2.529q_F$), the SCAN functional performs much worse than LDA, PBE, and PBEsol. This is somewhat surprising, because the SCAN functional is ostensibly designed to be superior over LDA and GGA functionals.
In contrast to that, LDA and PBEsol show a robust performance with an accuracy better than $4~\%$ at $q\leq 1.686~q_F$ for both $r_s=2$ and $r_s=6$.

As a key finding of our assessment, we highlight that the overall performance of the considered XC functionals deteriorates upon increasing the wave number of the density perturbation. 
At $r_s=2$ and $q= 2.529~q_F$, LDA and PBEsol still provide an accuracy with an error of less than $6~\%$ in $\delta n=n-n_0$. 
At $r_s=6$, the same is valid for PBEsol in the regime of weak perturbations, $\delta n/n_0\ll 1$. Other XC functionals essentially fail at $r_s=6$, $\delta n/n_0< 1$, and $q= 2.529~q_F$. At the largest considered wave number, $q= 5.9~q_F$, all considered XC functionals yield errors of less than $6~\%$ if $\delta n/n_0\ll 1$ for metallic densities ($r_s=2.0)$.
Finally, we highlight the failure of the considered XC functionals in the regime of strong perturbations, $\delta n/n_0>1$, and large wave-numbers, $q= 5.9~q_F$, where they exhibit a maximum deviation of about $10~\%$. We also showed that the reported errors in the density correlate with the errors in the total energy.

Based on the performed analysis we can formulate the following general recommendations for using XC functionals within the typical WDM regime at temperatures $T\leq T_F$ and when the total density is of interest:  When characteristic wave numbers $q<q_F$ are of interest, the SCAN functional is the most reliable choice, it provides accuracy at the level of QMC. For a wider range of wave numbers $q\leq 5.9~q_F$ (at $r_s=2$) and $q\leq 1.686~q_F$ (at $r_s=6$), the LDA and PBEsol functionals should be used if $\delta n/n_0<1$, because they provide consistent results with a relative error not exceeding a few percent. 

One can understand the performance of the considered XC functionals by recalling that, in the case of a weak perturbation, the density deviation from the mean value is defined by the static response function $\chi(q)$ since $\delta n(r)=2A\chi(\vec q)\cos{\vec q\cdot\vec r}$. Within linear response theory of the UEG, XC effects are included in terms of the local field correction $G(\vec q)$ via the relation $\chi(\vec q)^{-1}=\chi_0(\vec q)^{-1}-\frac{4\pi}{q^2}\left(1- G(\vec q)\right)$, where $\chi_0(\vec q)$ is the response function of ideal electron gas, i.e., the non-interacting UEG. In this case, the XC functional is $K_{\xc}(\vec q)=-\frac{4\pi}{q^2} G(\vec q)$. In the UEG, the LDA functional reduces to the long wavelength limit $q\ll 2q_F$ of $K_{\xc}$. Clearly, this behavior can be attributed to the failure of the LDA at perturbation wave numbers $q>2q_F$. The same is valid for PBE and PBEsol, where gradient corrections to exchange and correlation exactly cancel each other in the limit of the UEG. 
On the other hand, when the density perturbation is strong, the characteristic wave number of the density perturbation defined by the local gradient of the density is larger than the wave number of the external harmonic perturbation. In this regime, regions with a strong density localization are formed. These lead to an increase of the kinetic energy and reduce the importance the wave-number dependence in XC functionals. This may explain the better performance of the considered XC functionals when the perturbation amplitude is increased. 
Additionally, we note that the SCAN functional is designed to yield PBE-like results in the limit of the UEG. However, it is not trivial to explain the bad performance of SCAN compared to PBE in the case of a weak perturbation and $q>2q_F$ due to the large number of exact conditions enforced. A more in-depth analysis is needed to explain this behavior.

The presented data along with our assessment constitute an indispensable guide on the choice of the XC functional for KS-DFT calculations when an inhomogeneous electronic structure of WDM is investigated. This is of paramount importance for the diagnostics of XRTS experiments. We highlight the importance of this application by pointing out that KS-DFT results are used to extract electronic parameters like temperature and density from experimental observations. 
Besides that, our findings advance our understanding on how well KS-DFT is capable of capturing the electronic structure under WDM conditions. They also point to the parameter space where XC functionals ought to be improved for their use in the WDM application domain. 
We highlight the need for XC functionals that are accurate when perturbed electronic states are present. This goes beyond the inclusion of explicit temperature effects in the XC free energy.
This particular outcome of our assessment is valuable for DFT developers.

In the light of the vast amount of available XC functionals, a comprehensive assessment was beyond the scope of this work. 
An extensive comparison of available XC functionals shall be presented elsewhere. Instead, in this work, we focus on a representative set of XC functionals $-$ the basic LDA and its common generalizations. Furthermore, we have set up a numerical workflow for benchmarking XC functionals under perturbed electronic structures in WDM.
The presented KS-DFT data, input scripts, and QMC data will be made accessible online~\cite{data}. In doing so, we provide tools to test any existing or newly developed XC functional under perturbed electronic states in WDM. We believe that our work is a valuable contribution that facilitates both the rigorous assessment and construction of XC functionals adapted to the needs of WDM.

\section*{Acknowledgments}
We acknowledge helpful feedback from M. Bussmann. ZM gratefully acknowledges stimulating discussions with Timothy Callow.
This work was funded by the Center for Advanced Systems Understanding (CASUS) which is financed by the German Federal Ministry of Education and Research (BMBF) and by the Saxon Ministry for Science, Art, and Tourism (SMWK) with tax funds on the basis of the budget approved by the Saxon State Parliament. We gratefully acknowledge computation time at the Norddeutscher Verbund f\"ur Hoch- und H\"ochstleistungsrechnen (HLRN) under grant shp00026, and on the Bull Cluster at the Center for Information Services and High Performance Computing (ZIH) at Technische Universit\"at Dresden.

\section*{Appendix: KS-DFT Simulation Parameters}\label{sec:app} 
All KS-DFT calculations were performed with the GPAW code~\cite{GPAW1, GPAW2, ase-paper, ase-paper2}.
A \emph{k}-point grid of $N_k\times N_k\times N_k$ with $N_k=12$ at $r_s=2$ and $N_k=8$ at $r_s=6$ with a Monkhorst-Pack sampling of the Brillouin zone ($\vec k$-points) was used.
At $T=T_F$, 180 orbitals (with the smallest occupation number of about $10^{-4}$) were used for a total of $N=14$ electrons. The grid spacing was set to $0.15~{\rm \AA}$ for $0.02\leq A\leq 1$ and $r_s=2$, $0.05~{\rm \AA}$  for $A=5$ and $r_s=2$, $0.25~{\rm \AA}$ for $r_s=6$ and $0.05\leq A\leq 0.1$. At $A=5$, $240$ bands were used. Convergence criteria used for the self-consistency cycle: the energy change (last 3 iterations) must be less than $0.5~ {\rm meV}$ per valence electron, the change in integrated absolute value of density change must be less than 0.0001 electrons per valence electron, and the integrated value of the square of the residuals of the Kohn-Sham equations should be less than $4\times 10^{-8}~{\rm eV}^2$ per valence electron (see \href{https://wiki.fysik.dtu.dk/gpaw/documentation/manual.html#manual-convergence}{GPAW documentation}).

\begin{figure}
\center
\includegraphics{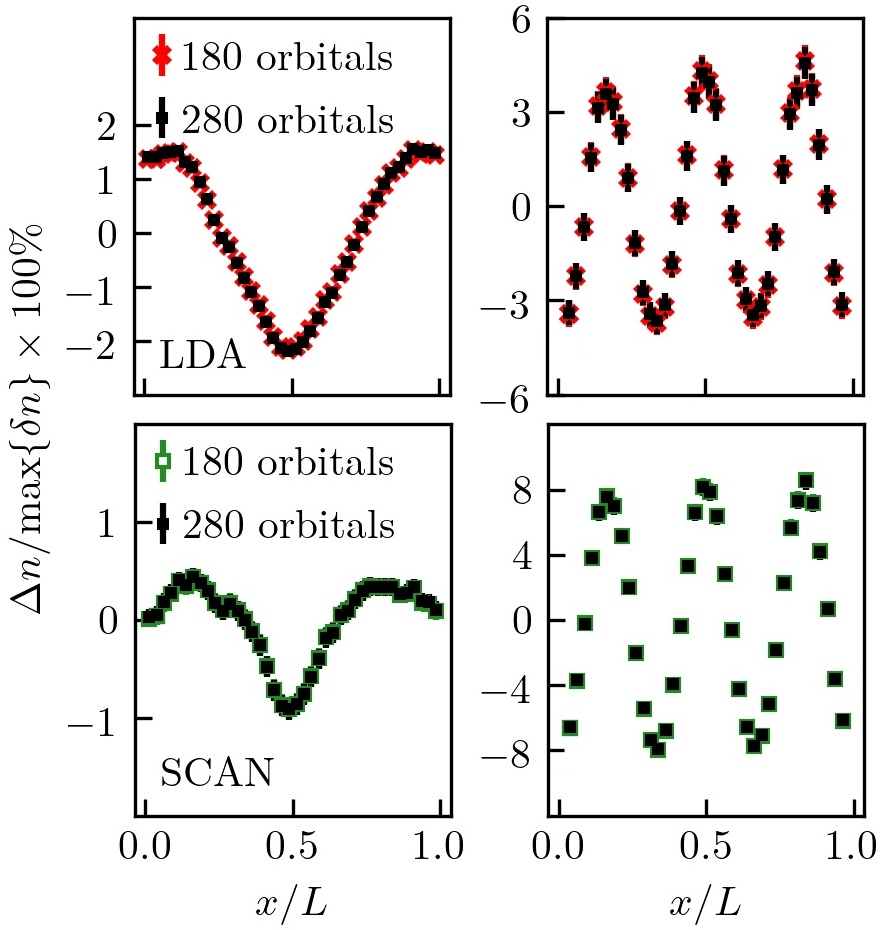}
\caption{\label{fig:Nband_test}
Relative deviation in the density $\Delta n/{\rm max }~\delta n  \times 100~\%$ between the KS-DFT data and the reference QMC data for $r_s=2$ with $A=0.1$. The wave numbers of the harmonic perturbation in Eq.~(\ref{eq:H}) are $q_{\rm min}$ (left) and $3 q_{\rm min}$ (right). Top panel: the data computed using LDA XC functional with $180$ and $280$ bands. Bottom panel: the data calculated using SCAN XC functional with $180$ and $280$ bands.}
\end{figure}

In Fig. \ref{fig:Nband_test} we compare results computed using $180$ bands and $280$ bands using LDA and SCAN. From Fig. \ref{fig:Nband_test} we see that there is no notable difference in the results for densities.

\begin{figure}
\center
\includegraphics{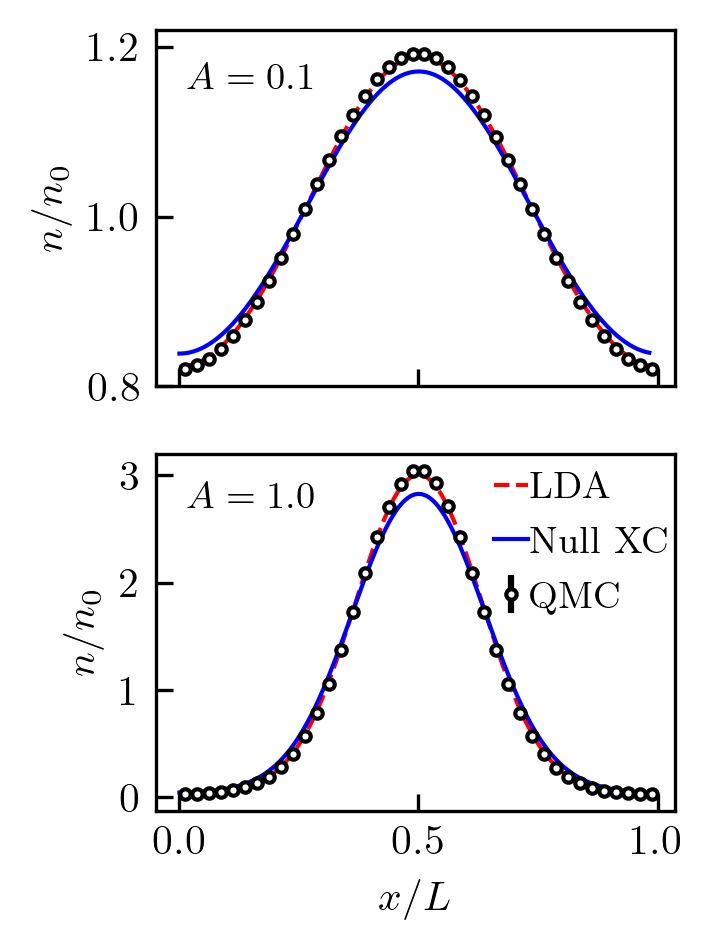}
\caption{\label{fig:NullXC}
The electronic density distribution along the perturbation direction for two different amplitudes $A$ at wave number $q_{\rm min}$ for $r_s=2$ and $T=T_F$.
The comparison of the data computed using LDA XC functional and null XC functional with QMC data shows the importance of the XC functional for the calculation of the electronic density.   
}
\end{figure}

In Fig. \ref{fig:NullXC}, we present results for the total density computed using null XC functional in comparison with the data obtained using QMC simulation and  KS-DFT simulation with LDA XC functional. From Fig. \ref{fig:NullXC} we see that it is essential to use XC functional to correctly compute the total density using KS-DFT method.

\section*{Appendix: PIMC simulation details}\label{sec:app2} 

The basic equation behind the PIMC method~\cite{cep,Berne_JCP_1982,Takahashi_Imada_PIMC_1984} is the canonical partition function
\begin{eqnarray}\label{eq:Z}
Z_{\beta,N,V} &=& \frac{1}{N^\uparrow! N^\downarrow!} \sum_{\sigma^\uparrow\in S_N} \sum_{\sigma^\downarrow\in S_N} \textnormal{sgn}(\sigma^\uparrow,\sigma^\downarrow)\\\nonumber & & \times \int d\mathbf{R} \bra{\mathbf{R}} e^{-\beta\hat H} \ket{\hat{\pi}_{\sigma^\uparrow}\hat{\pi}_{\sigma^\downarrow}\mathbf{R}}\ ,
\end{eqnarray}
where $\mathbf{R}=(\mathbf{r}_1,\dots,\mathbf{r}_N)^T$ contains the coordinates of all $N$ particles,  and $\hat{\pi}_{\sigma^\uparrow}$ ($\hat{\pi}_{\sigma^\downarrow}$) is the permutation operator corresponding to a particular element $\sigma^\uparrow$ ($\sigma^\downarrow$) from the permutation group $S_N$. In addition, $\textnormal{sgn}(\sigma^\uparrow,\sigma^\downarrow)$ is the sign function, which is positive (negative) for an even (odd) number of pair exchanges~\cite{Dornheim_permutation_cycles}.
Unfortunately, the matrix elements of the density operator $\hat\rho=e^{-\beta\hat{H}}$ cannot be evaluated in a straightforward way as the kinetic ($\hat{K}$) and potential ($\hat{V}$) contributions to the full Hamiltonian do not commute,
\begin{eqnarray}
\bra{\mathbf{R}}e^{-\beta\hat{H}}\ket{\mathbf{R}} \neq \bra{\mathbf{R}} e^{-\beta\hat{K}} e^{-\beta\hat{V}} \ket{\mathbf{R}}\ .
\end{eqnarray}
As a practical workaround, one uses the exact semi-group property of $\hat{\rho}$, which allows one to re-write Eq.~(\ref{eq:Z}) as
\begin{eqnarray}\label{eq:Z_modified}
Z_{\beta,N,V} &=& \frac{1}{N^\uparrow! N^\downarrow!} \sum_{\sigma^\uparrow\in S_N} \sum_{\sigma^\downarrow\in S_N} \textnormal{sgn}(\sigma^\uparrow,\sigma^\downarrow)\\\nonumber & & \times \int d\mathbf{R}_0\dots d\mathbf{R}_{P-1}
\bra{\mathbf{R}_0}e^{-\epsilon\hat H}\ket{\mathbf{R}_0}\\\nonumber & & \times \bra{\mathbf{R}_1}e^{-\epsilon\hat H}\ket{\mathbf{R}_1} \dots 
\bra{\mathbf{R}_{P-1}} e^{-\epsilon\hat H} \ket{\hat{\pi}_{\sigma^\uparrow}\hat{\pi}_{\sigma^\downarrow}\mathbf{R}_0}\ ,
\end{eqnarray}
where we have defined $\epsilon=\beta/P$. 
The comparison between Eqs.~(\ref{eq:Z}) and (\ref{eq:Z_modified}) reveals that the partition function has been transformed into a high-dimensional integration over $P$ density matrices that have to be evaluated at $P$ times the original temperature $T$. For a sufficiently large $P$, one can introduce a suitable high-temperature approximation like the simple \emph{primitive factorization} $e^{-\epsilon\hat{H}}\approx e^{-\epsilon\hat{V}}e^{-\epsilon\hat{K}}$. In fact, the associated factorization error decays as $P^{-2}$~\cite{brualla_JCP_2004,sakkos_JCP_2009}, and the convergence in the limit of $P\to\infty$ is ensured by the well-known Trotter formula~\cite{deRaedt_Trotter}. For completeness, we mention that higher-order factorizations of $\hat{\rho}$ are frequently employed in PIMC simulations~\cite{sakkos_JCP_2009,dornheim_njp15,dornheim_cpp_19}, although we do not find them necessary for the present study. In practice, we use $P=200$ primitive high-temperature factors such that any factorization error is substantially below the given level of statistical uncertainty; see the Supplementary Material of Ref.~\onlinecite{Dornheim_PRL_2020} for corresponding convergence plots.

The high dimensionality of Eq.~(\ref{eq:Z_modified}) requires a stochastic evaluation, which can be done efficiently using methods that are based on the celebrated Metropolis algorithm~\cite{metropolis}. Specifically, we employ a canonical adaption~\cite{mezza} of the worm algorithm by Boninsegni and co-workers~\cite{boninsegni1,boninsegni2}. Additional care has to be taken due to the antisymmetric nature of the fermionic density matrix, which can result in positive and negative contributions to Eq.~(\ref{eq:Z_modified}). The corresponding cancellation of positive and negative terms is the origin of the notorious fermion sign problem~\cite{dornheim_sign_problem,troyer}, which leads to an exponential increase in compute time with increasing the system size $N$ or decreasing the temperature $T$. 

In practice, this bottleneck is often avoided by imposing the \emph{fixed-node approximation}~\cite{Ceperley1991}. While this allows one to formally remove the sign problem, this advantage comes at the cost of an uncontrolled approximation~\cite{schoof_prl15}. Obviously, this is a disadvantage that severely limits the value of such data to benchmark other approximations like thermal DFT. Therefore, we do not impose any nodal restrictions in our simulations, and our PIMC results are exact within the given level of uncertainty. This is possible for the present parameters by using modern supercomputer clusters, and we have spend $\mathcal{O}(10^5)$ CPUh for the most costly data points with $N=34$ at $r_s=2$ and $\theta=1$.

\section*{Data Availability}
The data supporting the findings of this study are available on the Rossendorf Data Repository (RODARE)~\cite{data}.

\bibliography{ref, mb-ref}

\end{document}